\documentclass[iop,twocolappendix,numberedappendix]{emulateapj}
\usepackage{epstopdf}

\shortauthors{Ruan et al.}
\shorttitle{Variability in SEQUELS}
\begin{document}
\title{The Time-Domain Spectroscopic Survey: \\ Understanding the Optically Variable Sky with \emph{SEQUELS} in SDSS-III}
\author{John~J.~Ruan\altaffilmark{1,2}, 
Scott~F.~Anderson\altaffilmark{2},
Paul~J.~Green\altaffilmark{3}, 
Eric~Morganson\altaffilmark{3},
Michael~Eracleous\altaffilmark{4,5},
Adam~D.~Myers\altaffilmark{7},
Carles~Badenes\altaffilmark{8},
Matthew~A.~Bershady\altaffilmark{9},
William~N.~Brandt\altaffilmark{4,5}, 
Kenneth~C.~Chambers\altaffilmark{10},
James~R.~A.~Davenport\altaffilmark{2, 11, 12},
Kyle~S.~Dawson\altaffilmark{13},
Heather~Flewelling\altaffilmark{10},
Timothy~M.~Heckman\altaffilmark{14},
Jedidah~C.~Isler\altaffilmark{15},
Nick~Kaiser\altaffilmark{10},
Jean-Paul~Kneib\altaffilmark{16,17},
Chelsea~L.~MacLeod\altaffilmark{18},
Isabelle~Paris\altaffilmark{19},
Nicholas~P.~Ross\altaffilmark{18},
Jessie~C.~Runnoe\altaffilmark{4,5},
Edward~F.~Schlafly\altaffilmark{20},
Sarah~J.~Schmidt\altaffilmark{21,22},
Donald~P.~Schneider\altaffilmark{4,5},
Axel~D.~Schwope\altaffilmark{22},
Yue~Shen\altaffilmark{23,24},
Keivan~G.~Stassun\altaffilmark{15,25},
Paula~Szkody\altaffilmark{2},
Christoper~Z.~Waters\altaffilmark{10},
Donald~G.~York\altaffilmark{26}
}
\altaffiltext{1}{ Corresponding author: jruan@astro.washington.edu} 
\altaffiltext{2}{Department of Astronomy, University of
Washington, Box 351580, Seattle, WA 98195, USA}
\altaffiltext{3}{Harvard Smithsonian Center for Astrophysics, 
60 Garden St, Cambridge, MA 02138, USA}
\altaffiltext{4}{Department of Astronomy \& Astrophysics, 525 Davey Lab, The 
Pennsylvania State University, University Park, PA 16802, USA}
\altaffiltext{5}{Institute for Gravitation and the Cosmos, The Pennsylvania State 
University, University Park, PA 16802, USA}
\altaffiltext{6}{Department of Physics, 104 Davey Lab, The Pennsylvania
State University, University Park, PA 16802, USA}
\altaffiltext{7}{University of Wyoming, Dept. of Physics and Astronomy 3905, 1000 E. University, Laramie, WY 82071, USA}
\altaffiltext{8}{Department of Physics and Astronomy and Pittsburgh Particle Physics, Astrophysics, 
and Cosmology Center (PITT-PACC), University of Pittsburgh}
\altaffiltext{9}{Department of Astronomy, University of Wisconsin-Madison, 475 N. Charter
St., Madison, WI 53706, USA}
\altaffiltext{10}{Institute for Astronomy, University of Hawaii at Manoa, Honolulu, HI 96822, USA}
\altaffiltext{11}{Department of Physics \& Astronomy, Western Washington University, Bellingham, WA 98225}
\altaffiltext{12}{NSF Astronomy and Astrophysics Postdoctoral Fellow}
\altaffiltext{13}{Department of Physics and Astronomy, University of Utah, Salt Lake City, UT 84112, USA}
\altaffiltext{14}{Center for Astrophysical Sciences, Department of Physics and Astronomy, 
Johns Hopkins University, Baltimore, MD 21218}
\altaffiltext{15}{Vanderbilt University, Department of Physics \& Astronomy, Nashville, TN 37235, USA}
\altaffiltext{16}{Laboratoire d'astrophysique, Ecole Polytechnique F\'{e}d\'{e}rale
de Lausanne Observatoire de Sauverny, 1290 Versoix, Switzerland}
\altaffiltext{17}{Aix Marseille Universit\'{e}, CNRS, LAM (Laboratoire
d'Astrophysique de Marseille) UMR 7326, 13388, Marseille,
France}
\altaffiltext{18}{Institute for Astronomy, University of Edinburgh, Royal Observatory, 
Edinburgh, EH9 3HJ, United Kingdom}
\altaffiltext{19}{INAF - Osservatorio Astronomico di Trieste, Via G. B. Tiepolo 11, I-34131 Trieste, Italy}
\altaffiltext{20}{Max-Planck Institut f\"ur Astronomie, K\"onigstuhl 17, 69117,
Heidelberg, Germany}
\altaffiltext{21}{Department of Astronomy, Ohio State University, 140 West 18th Avenue,
Columbus, OH 43210}
\altaffiltext{22}{Leibniz-Institut für Astrophysik Potsdam (AIP), An der Sternwarte 
16, 14482, Potsdam, Germany}
\altaffiltext{23}{Kavli Institute for Astronomy and Astrophysics, Peking University,
Beijing 100871, China}
\altaffiltext{24}{Carnegie Observatories, 813 Santa Barbara Street, Pasadena, CA
91101, USA}
\altaffiltext{25}{Fisk University, Physics Department, Nashville, TN 37208, USA}
\altaffiltext{26}{Dept. of Astronomy and Astrophysics and the Enrico Fermi Institute, The University of
Chicago, 5640 South Ellis Avenue, Chicago, IL 60615, USA}


\begin{abstract}
The Time-Domain Spectroscopic Survey (TDSS) is an SDSS-IV eBOSS subproject primarily aimed at obtaining identification spectra of $\sim$220,000 optically-variable objects systematically selected from SDSS/Pan-STARRS1 multi-epoch imaging. We present a preview of the science enabled by TDSS, based on TDSS spectra taken over $\sim$320 deg$^{2}$ of sky as part of the SEQUELS survey in SDSS-III, which is in part a pilot survey for eBOSS in SDSS-IV. Using the 15,746 TDSS-selected single-epoch spectra of photometrically variable objects in SEQUELS, we determine the demographics of our variability-selected sample, and investigate the unique spectral characteristics inherent in samples selected by variability. We show that variability-based selection of quasars complements color-based selection by selecting additional redder quasars, and mitigates redshift biases to produce a smooth quasar redshift distribution over a wide range of redshifts. The resulting quasar sample contains systematically higher fractions of blazars and broad absorption line quasars than from color-selected samples. Similarly, we show that M-dwarfs in the TDSS-selected stellar sample have systematically higher chromospheric active fractions than the underlying M-dwarf population, based on their H$\alpha$ emission. TDSS also contains a large number of RR Lyrae and eclipsing binary stars with main-sequence colors, including a few composite-spectrum binaries. Finally, our visual inspection of TDSS spectra uncovers a significant number of peculiar spectra, and we highlight a few cases of these interesting objects. With a factor of $\sim$15 more spectra, the main TDSS survey in SDSS-IV will leverage the lessons learned from these early results for a variety of time-domain science applications.

\end{abstract}

\section{Introduction}
	The proliferation of large-scale time-domain imaging surveys has opened up a new window into the time-variable sky. Surveys including the Sloan Digital Sky Survey \citep[SDSS,][]{york00} Stripe 82 \citep{sesar07}, Pan-STARRS1 \citep[PS1,][]{kaiser02, kaiser10}, Catalina Sky Survey \citep[CSS,][]{drake09}, Palomar Transient Factory \citep[PTF,][]{law09}, La Silla-QUEST Variability Survey in the Southern Hemisphere \citep{hadjiyska12}, Lincoln Near-Earth Asteroid Research survey \citep[LINEAR,][]{stokes00}, OGLE I-OGLE IV surveys \citep{udalski08,wyrzykowski14}, Gaia \citep{gilmore12}, and future surveys such as the 
Zwicky Transient Facility \citep[ZTF,][]{bellm14} and the Large Synoptic Survey Telescope \citep[LSST, ][]{ivezic08} will provide publicly available light curves for up to billions of objects over large regions of sky. Many classes of objects exhibit broadband variability at optical wavelengths, and their diverse light curves contain rich temporal information to use in exploring their astrophysics.
	
	Spectroscopic follow-up of photometrically variable objects significantly enhances the science return on time-domain imaging surveys, by providing additional information on their physical nature and parameters that extends well beyond just object identifications/classifications and redshifts. For example, optical spectra of pulsating RR Lyrae stars discovered based on their periodic light curves provide information on their metallicities and radial velocities, enabling more accurate distance determinations and kinematic information. This is useful for probing the stellar distribution of the outer Galactic halo and discovery of halo substructures, which provides valuable insight on the formation of our Galaxy \citep[e.g.,][]{drake13a}. Furthermore, this spectral information is especially useful if large numbers of follow-up spectra are obtained in a systematic fashion, so that they can be mined for rare variable objects and events. For example, the elusive sub-parsec scale evolutionary phase of merging binary supermassive black holes may be observable through predicted signatures of periodic continuum emission in their broadband light curves \citep[e.g.,][]{dorazio13, farris15}, as well as offset broad emission lines in their optical spectra \citep[e.g.,][]{bogdanovic08, eracleous12}; evidence from a combination of both light curve information \citep[e.g.,][]{graham15, liu15} and spectra would result in a much stronger case for claims of their discovery. The large scope of current and future time-domain imaging surveys will yield of order $\gtrsim$10$^7$ time-variable objects across the sky, but spectroscopy on this scale will challenge spectroscopic resources and capabilities due to the sheer number of interesting variable objects that are expected to be discovered. To this end, a systematic spectroscopic survey dedicated to follow-up of variable sources will be highly complementary, innately full of scientific insights, and immensely valuable as a training set for time-domain science from current and future multi-epoch imaging surveys.

	The Time Domain Spectroscopic Survey (TDSS) is a large spectroscopic survey with the primary goal of obtaining a total of $\sim$220,000 single-epoch initial identification optical spectra of photometrically-variable objects (generally with $>$0.1 mag of variability) over 7,500 deg$^{2}$ of sky, selected based on the variability in their light curves from multi-epoch PS1 3$\pi$ survey (Chambers et al., in preparation) and SDSS imaging. As a subprogram of the extended Baryon Oscillation Spectroscopic Survey \citep[eBOSS;][]{dawson15} in SDSS-IV, these spectra will be obtained using the upgraded SDSS BOSS spectrograph over the full eBOSS footprint during the period of 2014-2020. TDSS targets are uniquely selected based on their flux variability amplitudes and neither on color nor detailed modeling of their light curve characteristics, providing a relatively inclusive spectroscopic sampling of the time-variable sky to complement various multi-epoch imaging surveys. Technical details of the survey plan and the target-selection method of the main TDSS program in SDSS-IV are described in \citet{morganson15}.
	
	In 2013-2014, SDSS-III undertook the Sloan Extended QUasar, Emission-Line galaxies, and Luminous red galaxies Survey (SEQUELS), an ancillary SDSS-III BOSS \citep{dawson13} dark-time program which is in part a pilot program for eBOSS in SDSS-IV. Since TDSS is a subprogram of eBOSS, spectra of TDSS targets were also obtained as part of SEQUELS. In this paper, we use the first 66 multi-object spectroscopic plates observed in the SEQUELS area to present early science results from these TDSS spectra, which will provide promising examples of the science enabled by TDSS in the full SDSS-IV survey. Specifically, we will first determine the demographics of our sample, which provides a ground-truth baseline for variability-selected samples since it is based on visual inspection of our spectra. We will then investigate the uniqueness and advantages of using variability-selected samples for various science applications, with the goal of informing detailed future studies on specific science questions using TDSS spectra.

	In Section 2, we describe the target-selection method of TDSS, the TDSS spectra taken as part of the SEQUELS survey, previous SDSS spectra of sample objects in this survey area, and our visual classification of these spectra. In Section 3,  we explore the general demographics and properties of these TDSS objects. In Section 4, we investigate the redshift distribution and spectral properties of quasars selected by TDSS, and discuss the unique characteristics inherent in variability-selected quasar samples. In Section 5 we investigate properties of variable stars in TDSS, including the chromospheric activity in M dwarfs, decomposition of spectroscopic binaries, and periodically variable stars. In Section 6, we investigate the possible origins of variability in a small sample of galaxies selected in TDSS. We summarize and conclude in Section 7.

\begin{figure*}[t]
\centering
\includegraphics[width=0.98\textwidth]{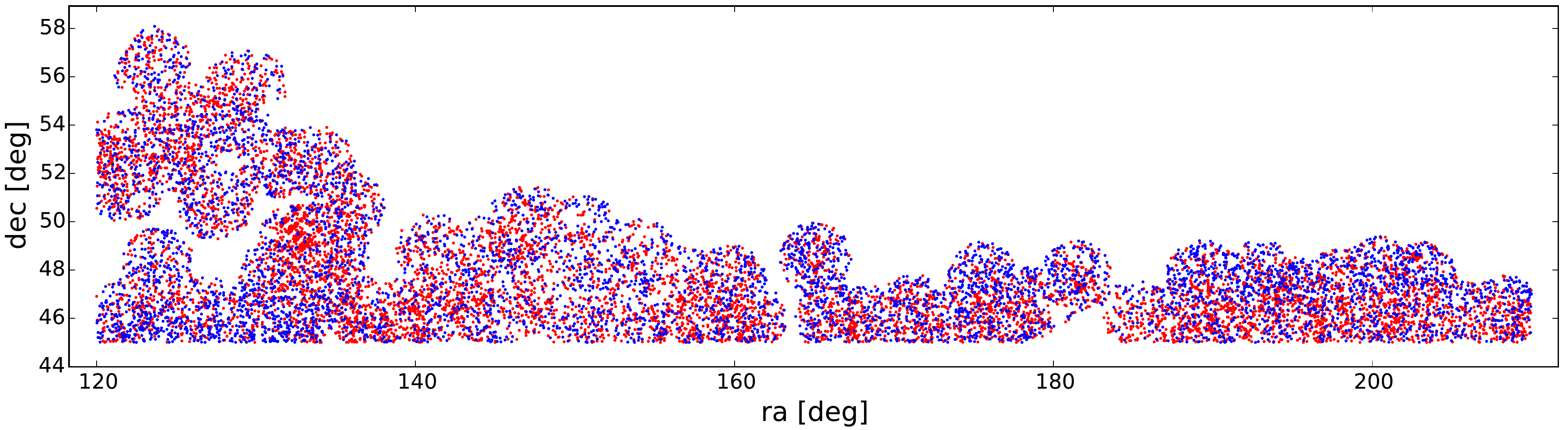}
\caption{The sky area of the 66 SDSS-III SEQUELS plates used in our investigation. Positions of newly-obtained spectra of TDSS-selected objects are shown as red points, and TDSS-selected objects with previous SDSS spectra are shown as blue points. The total geometric area of these plate areas is approximately 320 deg$^2$ (accounting for geometric plate overlaps but not detailed tiling of targets and adjacent plates, see Section 2.1).
}
\end{figure*}
	
\begin{deluxetable*}{cccc}
\tablecolumns{12}
\tablewidth{0pt}
\tablecaption{TDSS SEQUELS sample classifications from visual inspection of spectra}
\tablehead{\colhead{Object Type} &  \colhead{New SEQUELS spectra} & \colhead{Previous SDSS-I/II/III spectra}  & \colhead{Total spectra}}
\startdata
Quasars & 5503 & 4422 & 9925 \\
Stars & 4836 & 287 & 5123 \\
Galaxies & 292 & 30 & 322 \\
Other/Unknown & 343 & 33 & 376 \\
Total & 10,974 & 4,772 & 15,746
\enddata

\end{deluxetable*}

\begin{figure*}[t]
\centering
\includegraphics[width=0.49\textwidth]{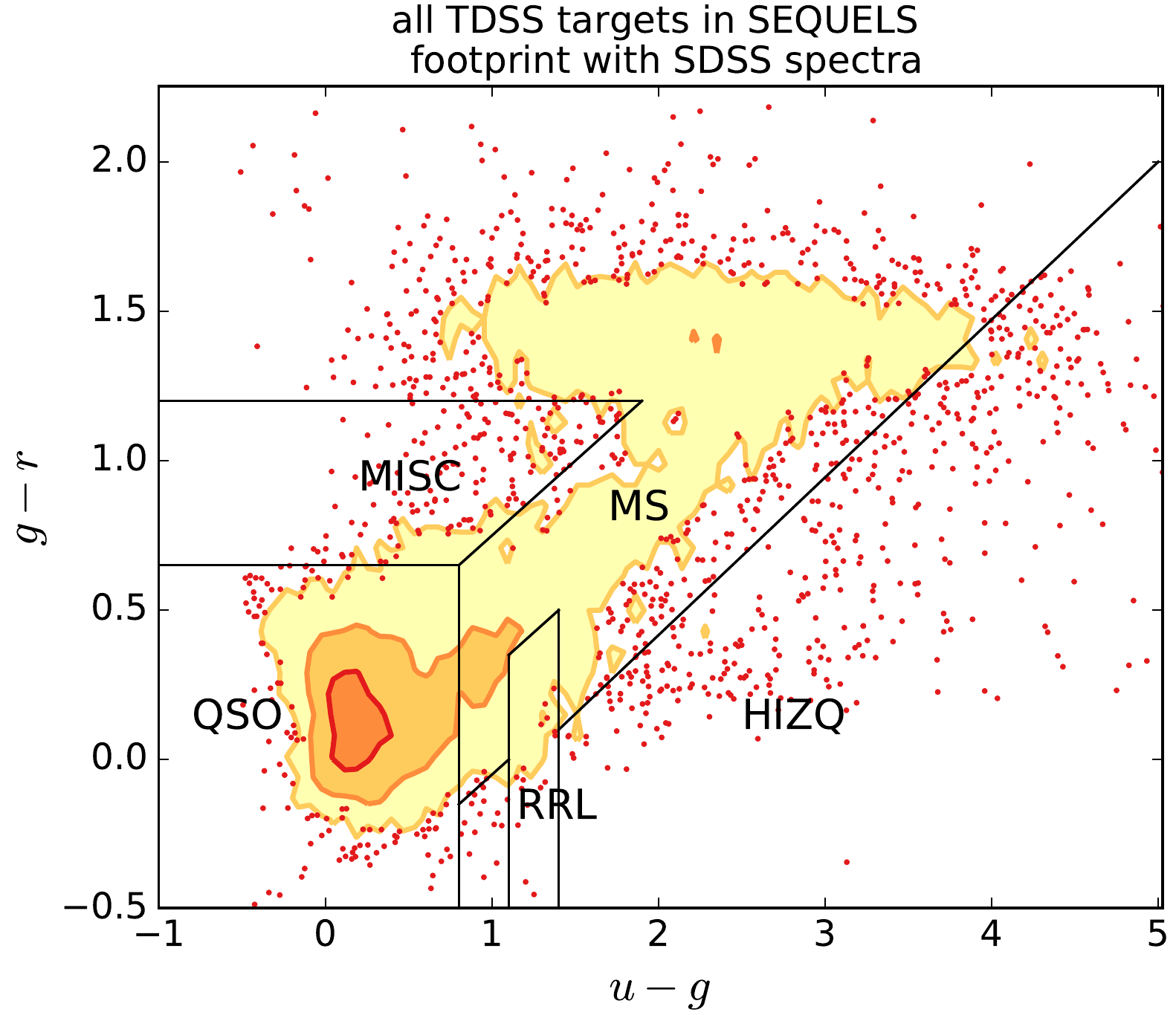}
\includegraphics[width=0.49\textwidth]{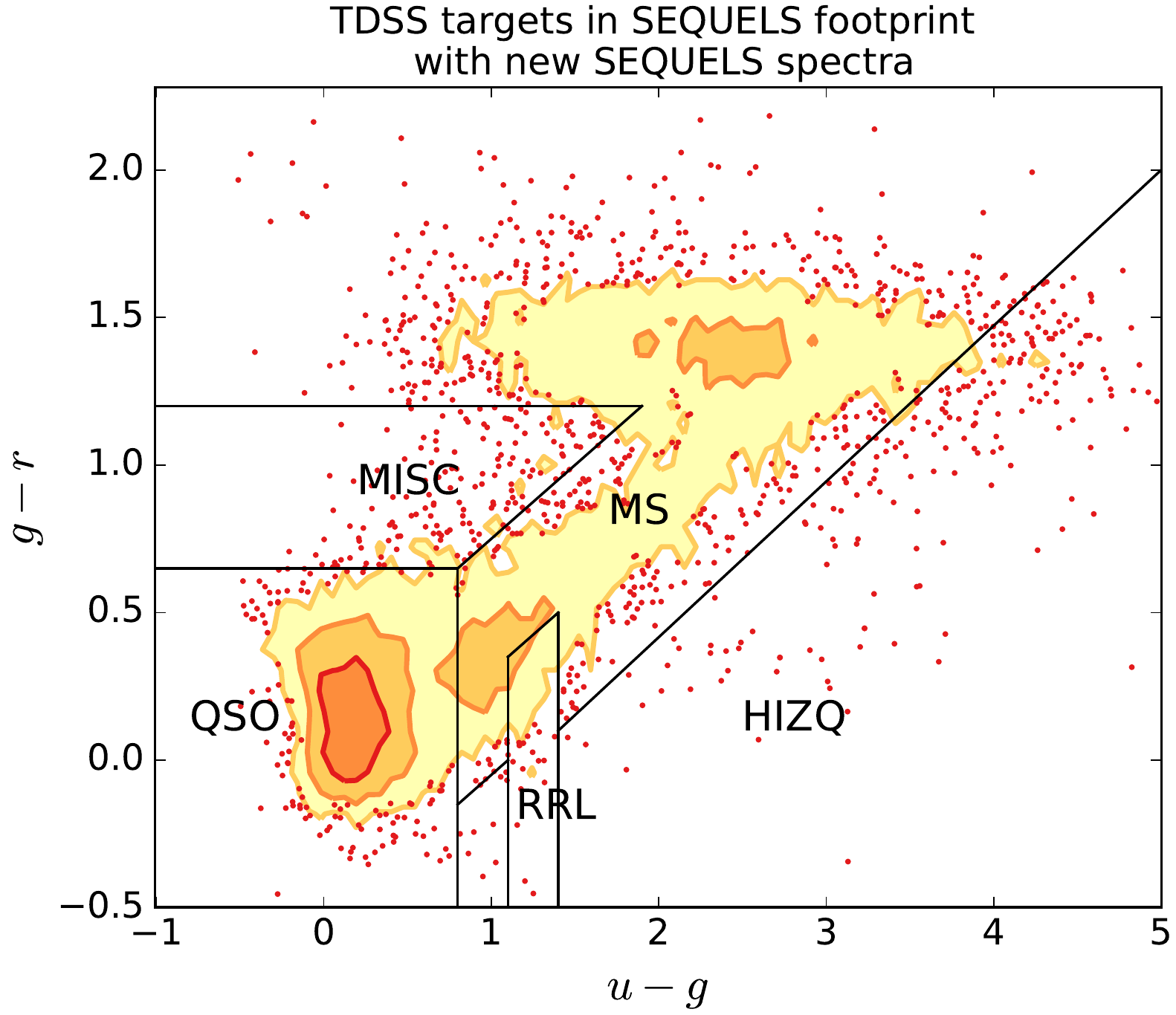}\\
\includegraphics[width=0.49\textwidth]{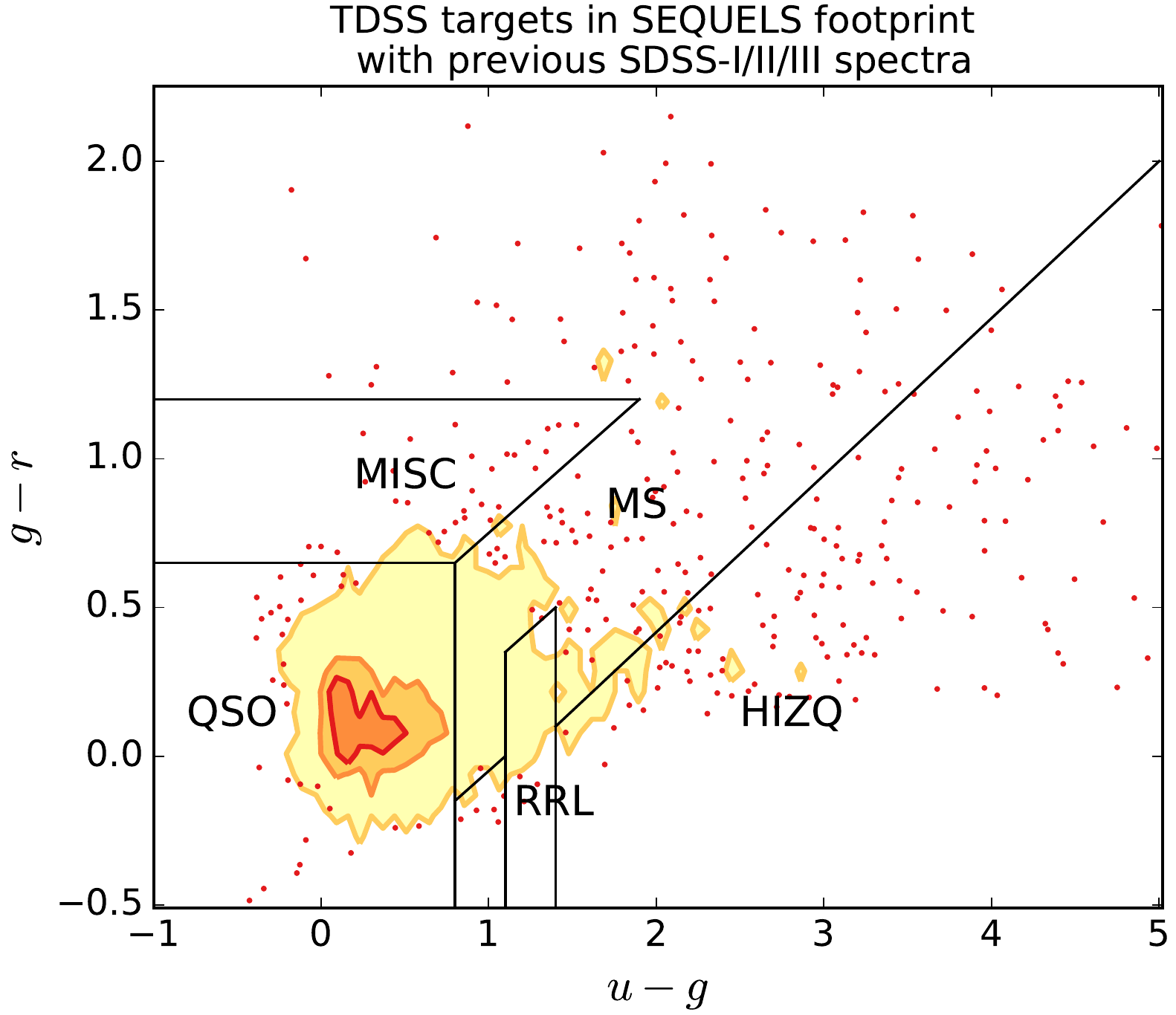}
\includegraphics[width=0.49\textwidth]{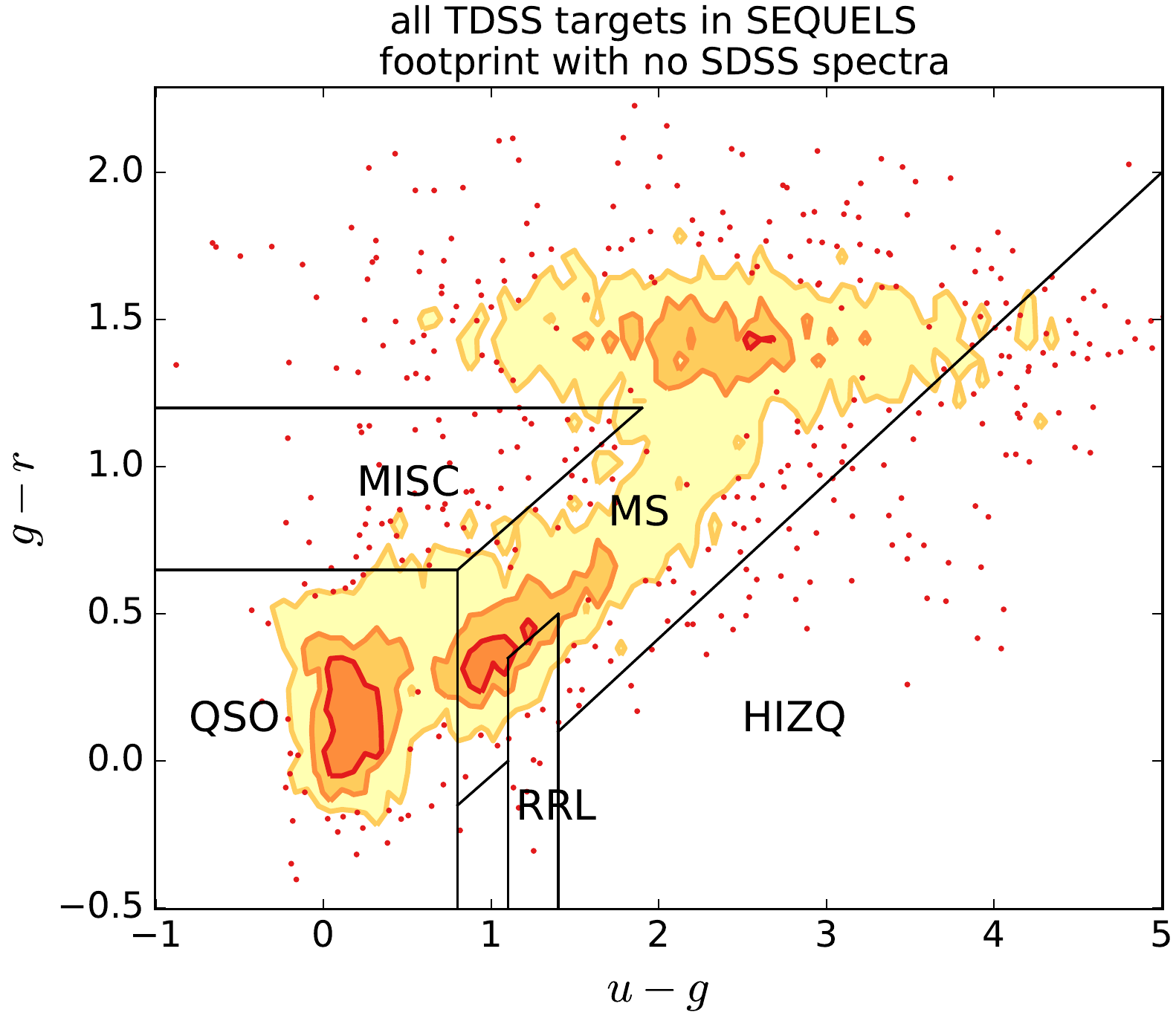}
\caption{Color-color diagrams (using SDSS photometry) of all 15,746 TDSS-selected objects with spectra in our SEQUELS sky area including previous spectra (top left), all 10,974 TDSS-selected objects with new spectra in SEQUELS (top right), all 4,772 TDSS-selected objects in our SEQUELS sky area with previous SDSS-I/II/III spectra (bottom left), and all 4,735 TDSS-selected objects in our SEQUELS sky area without spectra (bottom right). Contours enclose 20\%, 50\%, and 90\% of objects (from darkest to lightest), and the remaining 10\% are shown as red points. Regions in color-space containing mostly quasars, main-sequence stars, RR Lyrae, high-$z$ quasars, and other miscellaneous objects are labeled following the criteria in \citet{morganson15}. The sum of the objects in the top right and lower left panels gives the top left panel. These figures show that the vast majority of TDSS-selected objects with previous SDSS-I/II/III spectra have quasar-like colors, while the new SEQUELS spectra are a mix of quasars and stars. 
}
\end{figure*}
	
\section{Data}	
\subsection{TDSS Target Selection}	
	TDSS targets come from two types of observing programs: (1) a Single-Epoch Spectroscopy (SES) program encompassing $\sim$90\% of all TDSS spectra and aimed at obtaining initial single-epoch discovery spectra of photometrically variable objects, and (2) Few-Epoch Spectroscopy (FES) programs encompassing the remaining $\sim$10\% of TDSS spectra and consisting of various smaller projects involving multi-epoch spectroscopy of special targets. Targets for the FES programs are selected with a variety of methods to address more specific science questions. Since we are focusing on the general properties of variable objects in this current investigation, we utilize only SES spectra, which are instead selected specifically based on their variability properties in a systematic manner. Target selection for the TDSS SES spectra in the SDSS-III SEQUELS survey we use generally follows the method detailed in \citet{morganson15} for the main TDSS survey in SDSS-IV, but with minor deviations; below, we only summarize the most salient aspects and highlight the small differences in targeting between these two surveys.

	Photometric $griz$ light curves of point sources are constructed using a combination of SDSS Data Release 9 \citep[DR9,][]{ahn12} single-epoch imaging and PS1 3$\pi$ survey multi-epoch imaging. The TDSS targets are in essence selected by their (1) long-timescale ($\sim$10 years observed-frame) variability, (2) shorter-timescale ($\sim$2 years observed-frame) variability, and (3) median PS1 magnitudes. The long-timescale variability is measured as the difference in magnitude between the SDSS and median PS1 epoch magnitudes (where the SDSS magnitudes are color-corrected to match PS1 filters). The shorter-timescale variability is measured as the variance in the PS1 light curves. The median PS1 magnitudes are included in our target selection proceedure primarily to take into account the increase in required variability for fainter objects to enter our sample due to increased photometric uncertainties, and do not involve color information. These three parameters for each object are input into a 3-dimensional kernel density estimator (KDE) trained on the SDSS Stripe 82 variable object \citep{sesar07} and standard star (i.e., those that do not measurably vary) \citep{iverzic07} catalogs. The KDE assigns a probability that each object is variable, based on the efficiency $E$ with which variable objects are selected in its region of parameter space using the training sets. Specifically, $E = \rho_\mathrm{var} / \rho_\mathrm{std}$, where $\rho_\mathrm{var}$ and $\rho_\mathrm{std}$ are the KDE density of variable objects and standard stars from the training set, in the region of the 3-dimensional parameter space occupied by TDSS candidate variables. Candidate variable objects with $E$ above some threshold are then selected as `variable'. To ensure a uniform sky density of targets across the survey footprint, this $E$ threshold is calculated independently for different sky regions (each typically 4 deg$^2$) and is thus non-constant.
	
	The primary deviation of the TDSS targeting algorithm in SEQUELS from that of the main TDSS survey in SDSS-IV detailed in \citet{morganson15} is the sky density of TDSS-selected fibers. TDSS was allocated a fiber density budget of 20 deg$^{-2}$ in SEQUELS, double the density of 10 deg$^{-2}$ for the main TDSS survey in SDSS-IV. Since target selection is guided by these fiber-density constraints, the higher target fiber density of the TDSS sample in SEQUELS we use in this investigation generally leads to the inclusion of TDSS targets that would not meet the variability criteria to be selected in the same sky regions by the main TDSS survey in SDSS-IV. We note that SDSS plates are shared amongst several surveys, each with a number of differently-targeted samples that often contain overlaps. In particular, TDSS targets contain significant overlaps with the eBOSS CORE quasar sample, though the latter are intentionally color-selected to preferentially lie at redshifts of $z > 0.9$ \citep{myers15}. eBOSS aims to obtain a spectroscopic quasar sample (the eBOSS CORE sample) with high completeness and efficiency for cosmological quasar clustering studies over $0.9 < z <2.2$, and many of these quasars are also independently selected by TDSS due to their optical variability. Spectra of TDSS targets included in the eBOSS CORE sample were also obtained, but these fibers were not charged to the TDSS fiber density budget. These TDSS/CORE-selected spectra are included in our study, and we specifically investigate the differences between the TDSS and CORE quasar samples in Section 4.
	
	An additional difference in the TDSS targeting strategy between the SEQUELS sample and the main TDSS survey in SDSS-IV is that TDSS targets in SEQUELS were required to be detected in all four $griz$ bands, while this requirement was relaxed in SDSS-IV to allow inclusion of more high-redshift ($z>4$) quasars. Finally, unlike target selection in SDSS-IV TDSS, $\sim$30\% of the earliest SEQUELS TDSS targets were not visually pre-screened via imaging data before spectra were obtained. This visual pre-screening of TDSS targets uses cutouts of SDSS imaging to remove targets whose variability is suspect due to obvious photometry problems (e.g., lying on diffraction spikes of bright stars, lying within the isophotes of a spatially resolved galaxy, etc.). The majority ($\sim$75\%) of TDSS targets passed this visual prescreening, but the non-negligible $\sim$25\% which were rejected highlights the importance of diligent confirmation of variability in sparsely-sampled light curves from broadband imaging surveys. On the other hand, with a much higher surface density of potential targets than fibers allotted to TDSS in SDSS-IV, we have the luxury of being conservative in such exclusions and there are likely many genuine variables among this $\sim$25\%.

\subsection{SDSS-III SEQUELS Spectra}	
	Spectra of TDSS targets in SDSS-III  \citep{eisenstein11} were taken as part of the SEQUELS survey, and are publicly available\footnote{https://www.sdss3.org/} in the SDSS-III Data Release 12 and described in \citet{alam15}. These spectra were taken with the SDSS BOSS spectrograph \citep{smee13} on the SDSS 2.5m telescope \citep{gunn06}, with spectral resolution of $R\sim2000$ and wavelength coverage from approximately 3600\AA~to 10,400\AA, reduced with the SDSS-III BOSS spectral reduction pipeline \citep{bolton12}. 
	
	The SEQUELS spectra we utilize are from a set of 66 plates observed and reduced by June 15, 2014, and the sky coverage of these plates is shown in Figure 1. This does not encompass the full SEQUELS survey, and many more SEQUELS plates were observed after this date. We estimate the combined geometric plate area of this portion of the survey to be $\sim$320 deg$^{2}$ by Monte Carlo simulation, which takes into account plate overlaps. However, we caution that this plate area is approximate and in some ways misleading; some sky regions observed with one SEQUELS plate will not have complete spectroscopic coverage of all targets due to other plates overlapping the same region that had not yet been observed as part of the the survey tiling plan. In the sky area of these 66 SEQUELS plates, TDSS selects 20,184 variable objects for single-epoch spectroscopy; spectra were obtained for 15,746 (about 78\%) of these objects (through mid-June 2014). This sample of 15,746 TDSS spectra we use in this study consists of two parts: (1) 10,974 TDSS-selected objects with new spectra taken as part of SEQUELS, and (2) 4,772 TDSS-selected objects in our SEQUELS sky area with previous SDSS-I/II/III spectra. These two subsamples are mutually exclusive since TDSS purposely avoided targeting objects with previous SDSS spectra. To create the sample of TDSS targets with previous SDSS spectroscopy, we match all SDSS-I/II spectra from the SDSS spectrograph and SDSS-III spectra from the BOSS spectrograph (excluding SEQUELS TDSS targets) to the plate areas of the 66 SEQUELS plates we use in this study. These single-epoch TDSS spectra in SEQUELS can be selected from the SDSS DR12 databases using the {\tt EBOSS\_TARGET0} selection flag \citep[for details on various TDSS targeting flags in SEQUELS, see][]{alam15}.

\subsection{Visual Classification of Spectra}	
	The vast discovery space in a large-scale time-domain survey has the potential to uncover large numbers of peculiar objects. To ensure that such serendipitous discoveries are not missed, we visually inspect the SDSS spectrum of each of the 15,746 TDSS-selected variable objects in the SEQUELS (including those with previous SDSS spectra in our SEQUELS sky area). Although the SDSS-III pipeline \citep{bolton12} provides automated classification and redshift estimation which generally guides the visual inspection, we find that inspection of each object is crucial in identifying rare cases where the pipeline fails, and to flag especially peculiar objects. In the visual inspection, we classify and group these TDSS-selected objects into quasar (9,925), stellar (5,123), galaxy (322), and other/unknown (376) subsamples. Out of the other/unknown objects subsample, 32 objects had genuinely interesting spectra that were not classifiable (and not simply due to low signal-to-noise), 109 objects had a blank spectrum (e.g. due to dropped fibers), 30 objects had obvious glitches in the spectrum, and the remaining 205 had low signal-to-noise ratios that made classification difficult. These subsamples are summarized in Table 1, and used in our investigation of the properties of these different objects.
	
\begin{figure}[t]
\centering
\includegraphics[width=0.49\textwidth]{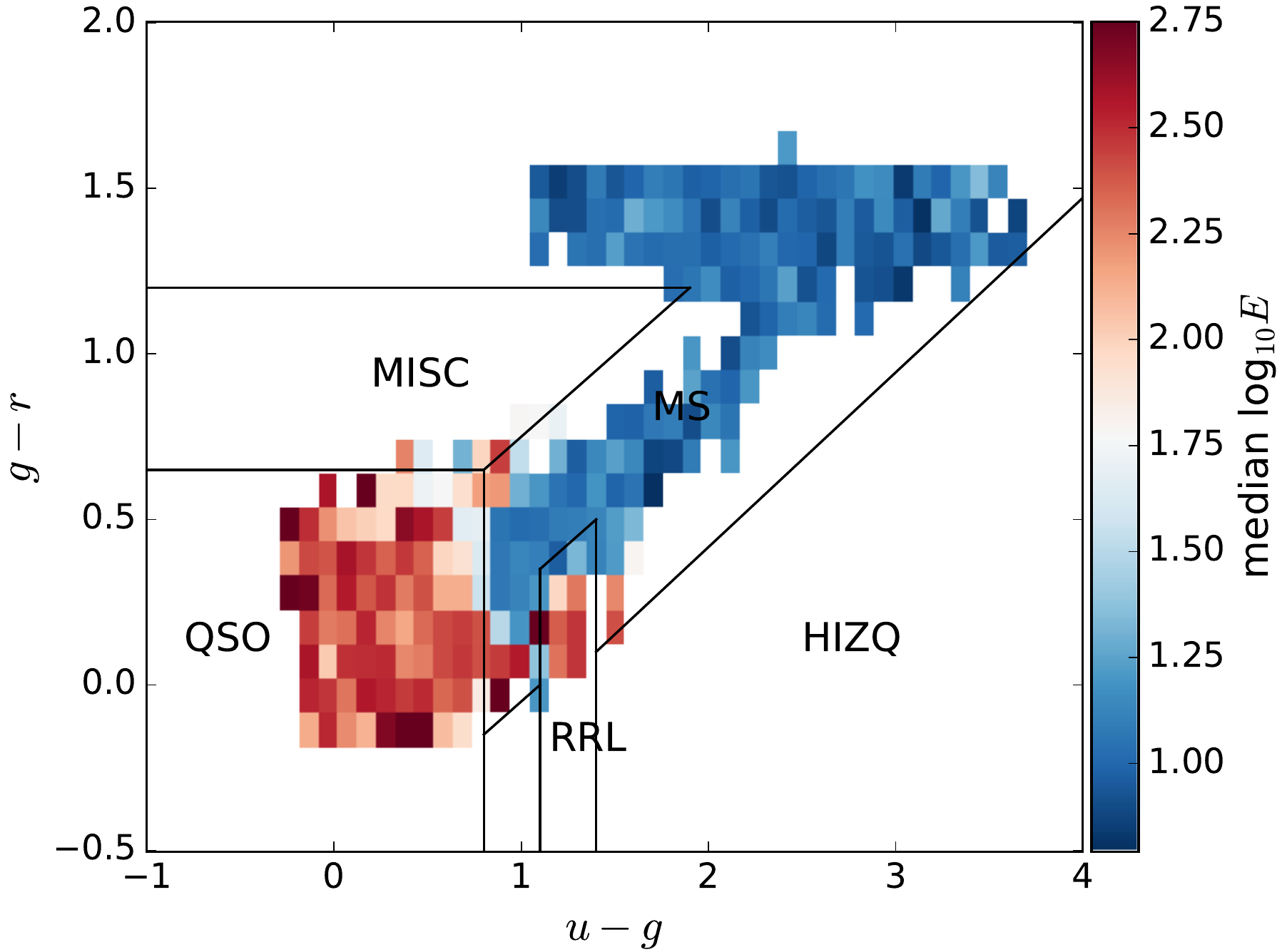}
\caption{Color-color diagram of the variability amplitude $E$ parameter defined in the \citet{morganson15} TDSS targeting paper, for all TDSS-selected objects with spectra in our SEQUELS sky area. Quasars and RR Lyrae stars appear to be the most variable (large $E$) object classes selected by TDSS, as expected.
}
\end{figure}

\section{General Properties of the TDSS Variability-Selected Sample}

	Recall that TDSS by design selects targets solely on the basis of optical variability, and in particular does not invoke color cuts or any other criteria that might intentionally select specific classes of astrophysical objects. Therefore, to assess the general properties of TDSS-selected objects in SEQUELS, we show in Figure 2 (top left) a color-color diagram of all 15,746 TDSS targets with spectra in our SEQUELS area. Since TDSS intentionally avoided targets in SEQUELS with previous SDSS spectra (except for the relatively small TDSS Few-Epoch Spectroscopy programs which are not included in this investigation), these 15,746 TDSS targets with spectra in our SEQUELS sky area are essentially the sum of the 10,974 TDSS targets with new SEQUELS spectra (top right) and the 4,772 TDSS targets with previous SDSS spectra (bottom left) also shown in Figure 2. The $u-g$ and $g-r$ colors for each object in Figure 2 and all other diagrams are derived from PSF magnitudes from SDSS photometry, in the SDSS filters (which are slightly different from the overlapping PS1 $grizy$ filters). As expected, the majority of TDSS targets with spectra in our SEQUELS sky area have quasar-like colors, and the subset with previous SDSS spectra are almost entirely (93\%) quasars. However, a significant number of objects are found in the stellar locus since the uniquely inclusive variability selection criteria of TDSS also select a wide variety of variable stars.
	
\begin{figure*}[t]
\centering
\includegraphics[width=0.49\textwidth]{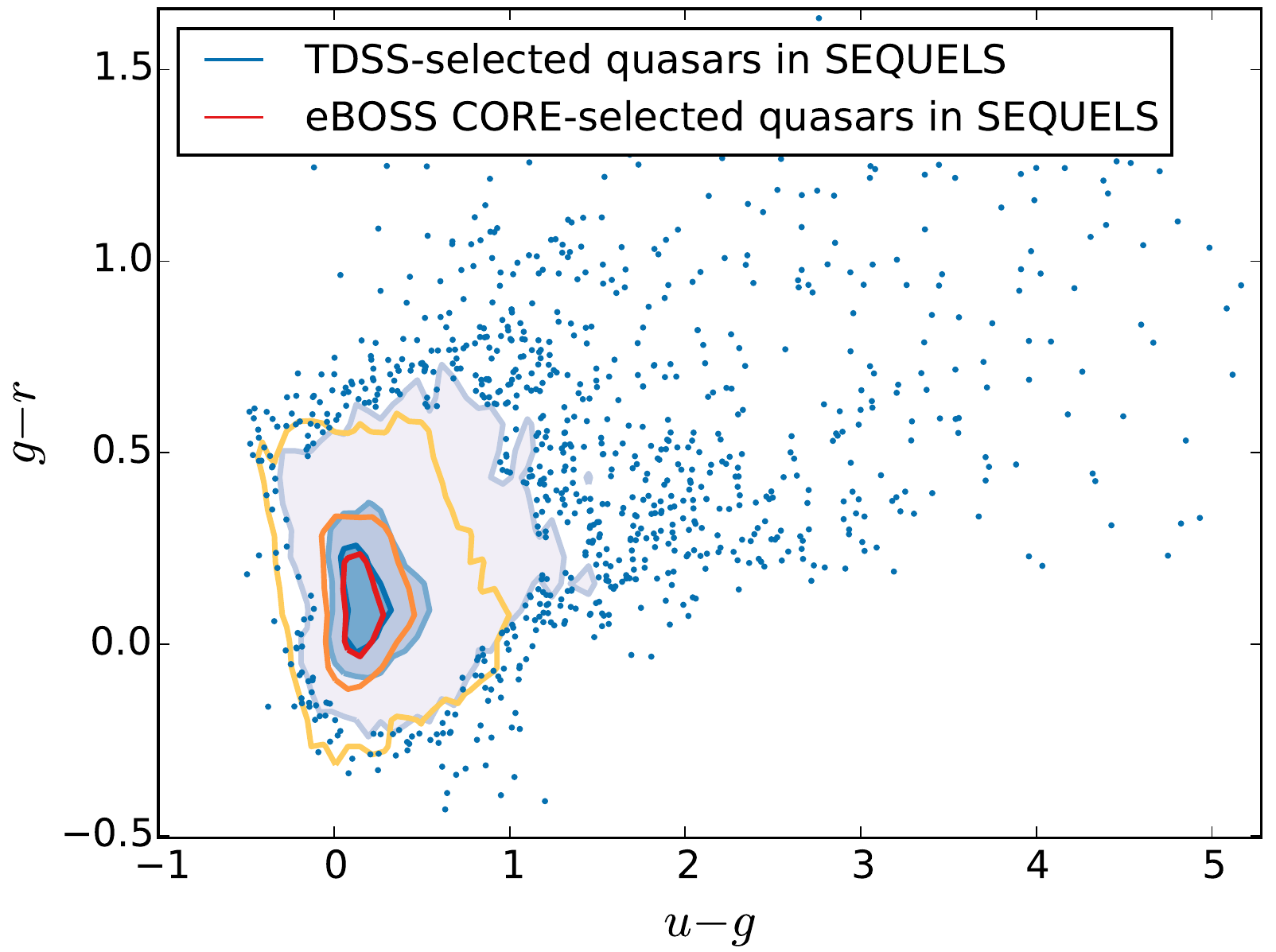}
\includegraphics[width=0.49\textwidth]{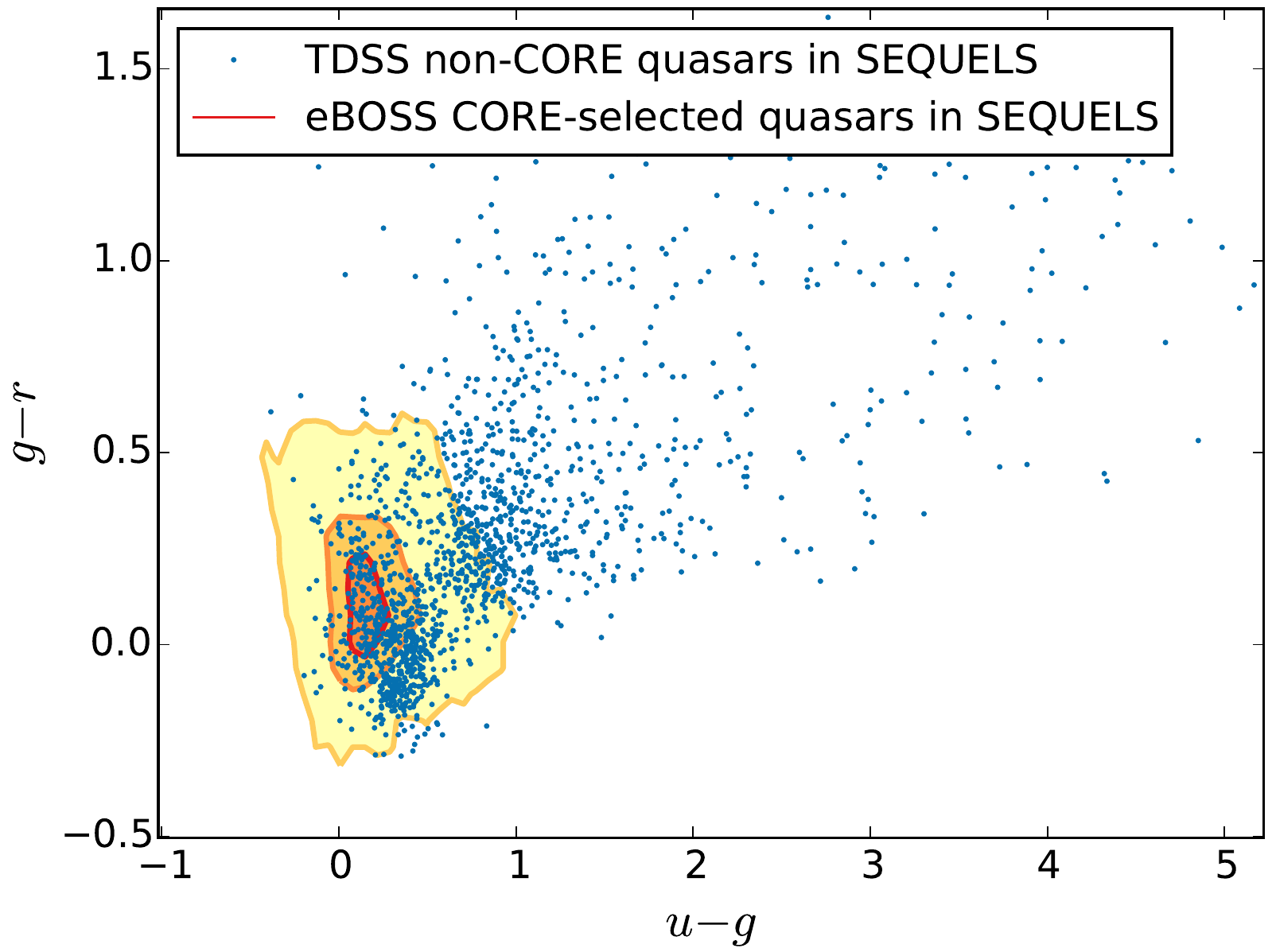}
\caption{Left: Color-color diagram of TDSS-selected quasars (blue contours), and eBOSS CORE quasars (red contours), with new SEQUELS spectra. Contours enclose 20\%, 50\%, and 90\% of the objects in each sample (from darker to lighter). The remaining 10\% of objects outside of the contours for the TDSS-selected sample are plotted as blue points. There is significant overlap between these two samples, but TDSS selects an additional population of redder quasars than the CORE sample. Right: eBOSS CORE quasars with new SEQUELS spectra (red contours, same as the red contours in left panel), and TDSS-selected quasars which were not selected by the CORE sample (blue points). 
}
\end{figure*}

	In the SEQUELS sky area of the 66 plates we use, there are an additional 4,735 TDSS targets for which no previous SDSS spectra exist and no new SEQUELS spectra were obtained through mid-June 2014, but this is primarily due to fiber allocation limits on the TDSS survey. Figure 2 (bottom right), shows a color-color diagram of these 4,735 TDSS targets that do not yet have spectra; this population of objects does not visually appear to be significantly different than the sample with new SEQUELS spectra (top right). This confirms that the TDSS-selected sample with spectra that we use in this study is similar to to these TDSS-selected objects without spectra. More importantly, this suggests that our conclusions on quasar and stellar properties are generally robust to survey incompleteness (e.g., the quasar redshift distribution should not significantly change with a more complete TDSS spectral sample).
	
	The variables selection efficiency parameter $E$  utilized in TDSS target selection can be thought of as an indirect measure of variability amplitude, since a large $E$ for a object implies its has a high probability of being a variable, due to its large variability amplitude. Figure 3 shows a color-color diagram of the $E$ parameter calculated for all TDSS-selected objects with spectra in our SEQUELS sky area as part of our KDE target selection procedure. Bluer objects with quasar-like colors have significantly larger $E$ than the redder objects along the stellar locus, although the region of color-space occupied by RR Lyrae has $E$ similar to quasars. This is qualitatively consistent with results from the analysis of SDSS Stripe 82 light curves by \citet{sesar07}.

\section{Quasars in TDSS} 
	Quasars are well-known to have highly variable continuum emission in the optical and UV, displaying $\sim$10-30$\%$ variability in flux over long timescales. Detailed modeling of quasar light curves has shown that the broadband variability is well-described to first approximation by a first-order continuous autoregressive process \citep[i.e.\ a damped random walk, e.g.][]{kelly09, kelly11, macleod10}, although small deviations from this behavior have been observed at short timescales using well-sampled light curves \citep[e.g.][]{mushotzky2011, zu13, edelson2014, graham14}. This variable quasar continuum emission is believed to originate from their accretion disks, but the cause of the long-term variability is unclear and has been suggested to originate from localized thermal fluctuations in the disks \citep{kelly09, dexter09, schmidt12, ruan14a}. Intriguingly, this optical variability is known to scale with quasar properties such as black hole mass and bolometric luminosity as measured from their optical spectra \citep{vandenberk04, wilhite08, macleod10}. Large samples of quasar light curves from imaging surveys can also be used to study spectral time-lags in the continuum emission \citep[][]{edelson2015}, and perform multi-object reverberation mapping of broad line regions \citep[e.g.][]{shen15a, shen15b}. In these and other science goals, complementary TDSS spectroscopy of these variable quasars provides crucial spectral information on their physical properties.
	
	\citet{morganson15} showed that the majority of variable objects selected by TDSS have blue, quasar-like colors, although variability-based selection is known to also efficiently select a significant number of quasars with redder, stellar-like colors in comparison to color-based selection \citep{butler11, macleod11, palanque11}. We directly compare the colors of variability-selected quasars in TDSS to color-selected quasars in the eBOSS CORE sample in SEQUELS (which uses the XDQSO method, \citealt{bovy12}), to understand the differences in the quasar populations selected by these two methods. The eBOSS CORE sample is primarily designed to yield a complete and efficient color-selected sample in the redshift range of $0.9 < z < 2.2$ for clustering measurements \citep{myers15}, although the targeting algorithm is free to select higher-$z$ ($z > 2.2$) quasars. Despite the vastly different approaches in target selection for the TDSS and eBOSS CORE quasar samples in SEQUELS, the fibers are interspersed amongst the same 66 plates, and thus there are no differences in the observing procedure. 
	
	Figure 4 (left) shows a color-color diagram of TDSS-selected quasars in our SEQUELS plate area, as well as those selected by the eBOSS CORE quasar algorithm (including those with previous SDSS-I/II/III spectra). As expected, there are strong overlaps between these two samples, but the TDSS-selected sample extends further into redder colors. To better highlight these TDSS-selected quasars not selected by the eBOSS CORE quasar sample, Figure 4 (right) shows these TDSS non-CORE quasars in SEQUELS, plotted over the same eBOSS CORE quasar sample contours as the left panel. Again this shows a population of red quasars extending into the stellar locus ($u-g \gtrsim 0.6$) which are not selected by the color-based eBOSS CORE quasar selection, but also a clump of blue quasars centered at approximately $g-r$, $u-g$ = [0.0, 0.4]. The redder quasars selected by variability but not color-selection can have redder colors because they may preferentially extend to higher redshifts, or they may be intrinsically redder (e.g. due to higher levels of host-galaxy dust extinction). We explore each of these two possibilities in turn below, and show that the redder TDSS quasars are in fact due to a combination of both these effects. Furthermore, we will show that the blue clump of quasars selected by variability but not colors in Figure 4 (right) are primarily low-redshift ($z<0.9$) quasars outside of the targeted redshift range of the eBOSS CORE sample.
	
\begin{figure}[t]
\centering
\includegraphics[width=0.47\textwidth]{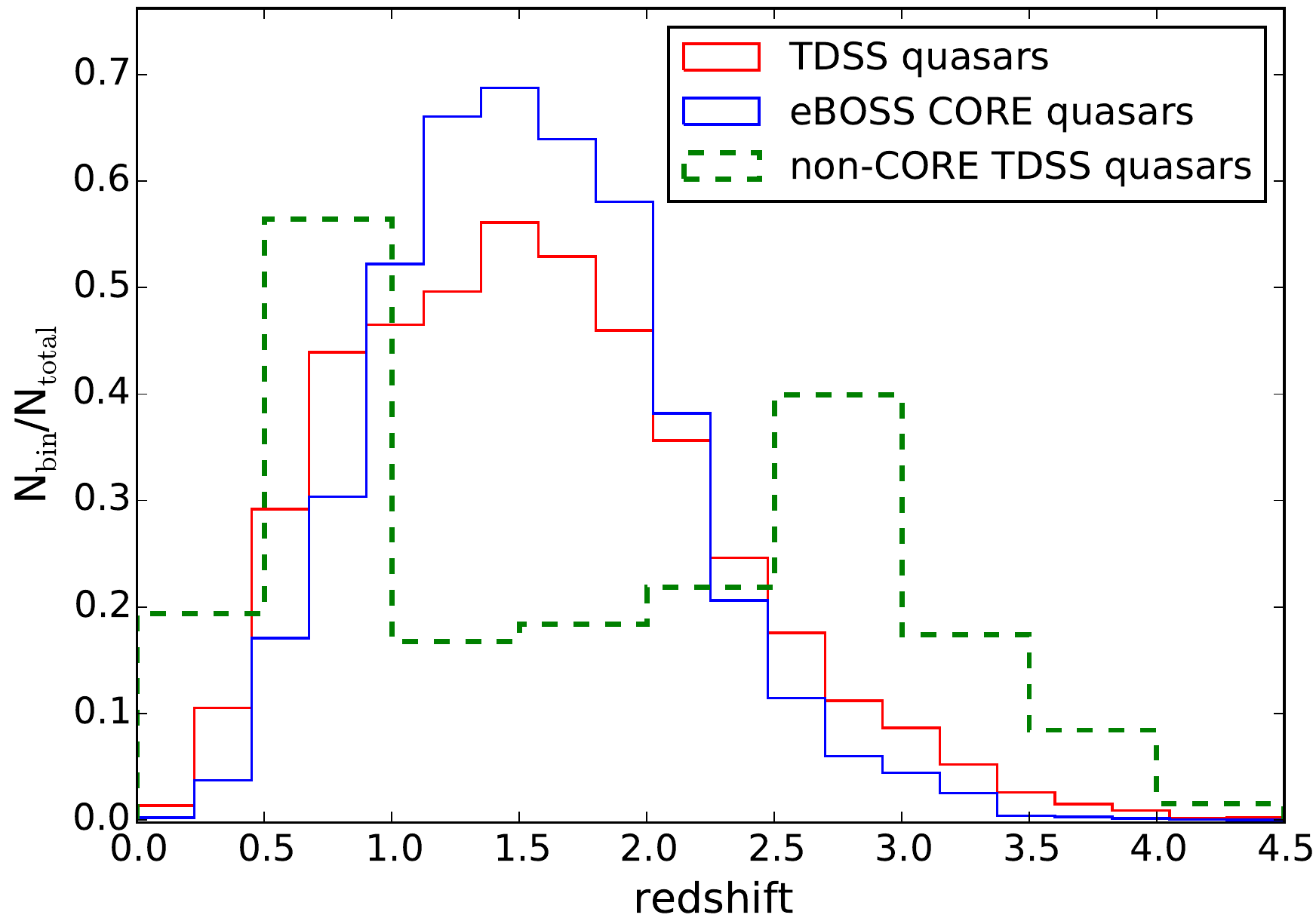}
\includegraphics[width=0.47\textwidth]{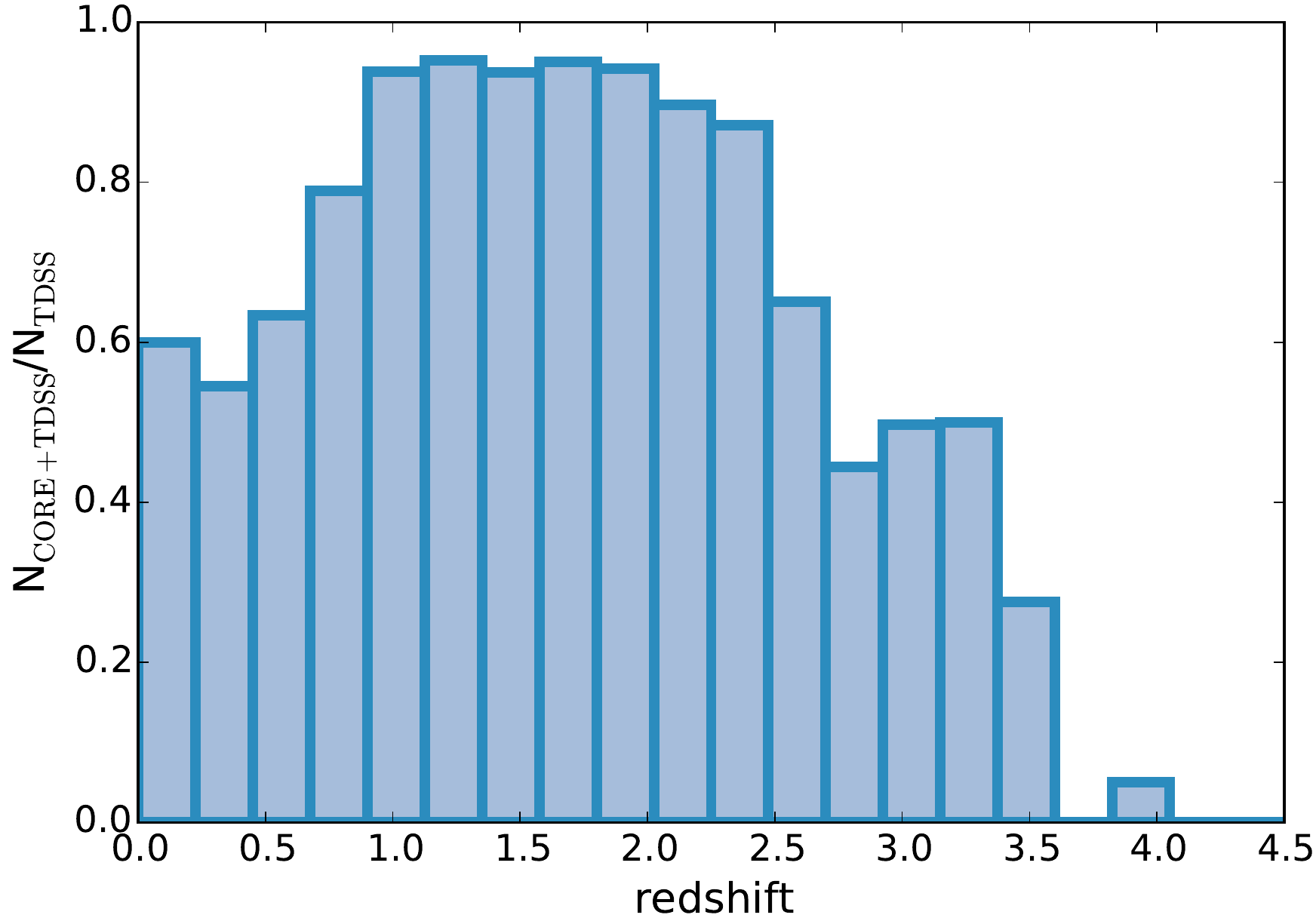}
\caption{ Top: Normalized redshift distribution of quasars in our SEQUELS sky area selected by TDSS (red solid), eBOSS CORE (blue solid), and the subset of TDSS quasars not selected by the CORE sample (green dashed). The inclusive variability-based approach of TDSS selects quasars with a broad, smooth redshift distribution. Bottom: Fraction of TDSS-selected quasars that are also selected by the eBOSS CORE algorithm, as a function of redshift. 
}
\end{figure}
	
\subsection{TDSS Quasar Redshift Distribution}
	Color-based selection of quasars has traditionally been a popular method to apply to large-scale imaging surveys extending into SDSS \citep[e.g.,][]{richards02, ross12}. The resulting large samples of quasars are useful for cosmological studies of the Ly$\alpha$ forest \citep{slosar13, busca13, fontribera14}, along with its various cross-correlations \citep[e.g.][]{fontribera12, shen13}, the quasar luminosity function \citep[e.g.,][]{ross13, mcgreer13}, quasar clustering \citep[e.g.][]{myers06, myers07a, myers07b, ross09}, and quasar physics \citep[e.g.,][]{filiz13, hall13, cai14}. In many of these science goals, high efficiency of selection and high sample completeness are required over large areas of sky. However, it is well-known that color-based selection can be inefficient and lead to incomplete samples in some regimes; for example, at higher redshifts ($z\gtrsim2$), quasar colors intersect the stellar locus in optical color-space, leading to a dearth of quasars in the $2.2 \lesssim z \lesssim 3$ redshift range \citep{fan99}. This problem is exacerbated by photometric uncertainties, which further blend together the quasar and stellar populations in color space, although it has been demonstrated that probabilistic approaches to deconvolution can lead to improvements in selection efficiency \citep{bovy11}.
	
	Unlike color-selection, a variability-selected sample of quasars is known to suffer less from contamination issues with the stellar locus \citep[][]{schmidt10, macleod11}. Since nearly all Type 1 quasars are variable in the optical \citep{sesar07}, our variability-selected sample is likely to be relatively complete and display a redshift distribution relatively close to the intrinsic Type 1 quasar redshift distribution. To understand the differences between our variability-selected quasar sample and a color-selected sample, we compare the redshift distribution of TDSS quasars to the eBOSS CORE sample. Figure 5 (top) compares the redshift distribution of TDSS-selected spectroscopic quasars in our SEQUELS plate area to those selected by the color-based eBOSS CORE quasar algorithm, using normalized histograms. These distributions for the TDSS and eBOSS CORE samples include quasars in our SEQUELS plate area that have previous SDSS spectra. Not surprisingly, the TDSS-selected sample shows a broader redshift distribution, selecting quasars at both higher ($z\gtrsim2.2$) and lower ($z\lesssim1$) redshifts, outside of the redshift range primarily targeted by the eBOSS CORE algorithm. Thanks to the relatively unbiased nature of variability selection, the redshift distribution of the TDSS-selected quasars in Figure 5 (top) is smooth and should be close to the intrinsic redshift distribution of quasars (insofar as the redshift distribution does not strongly depend on variability amplitude, although cosmological time-dilation may affect selection of high-$z$ quasars depending on their exact variability characteristics). Although the TDSS quasar sample used in this redshift distribution is not complete since there are TDSS-selected quasars in our SEQUELS plate area with post-DR12 spectra, the addition of these missing quasars is unlikely to introduce significant changes to this smooth redshift distribution.
 	
\begin{figure}[t]
\centering
\includegraphics[width=0.47\textwidth]{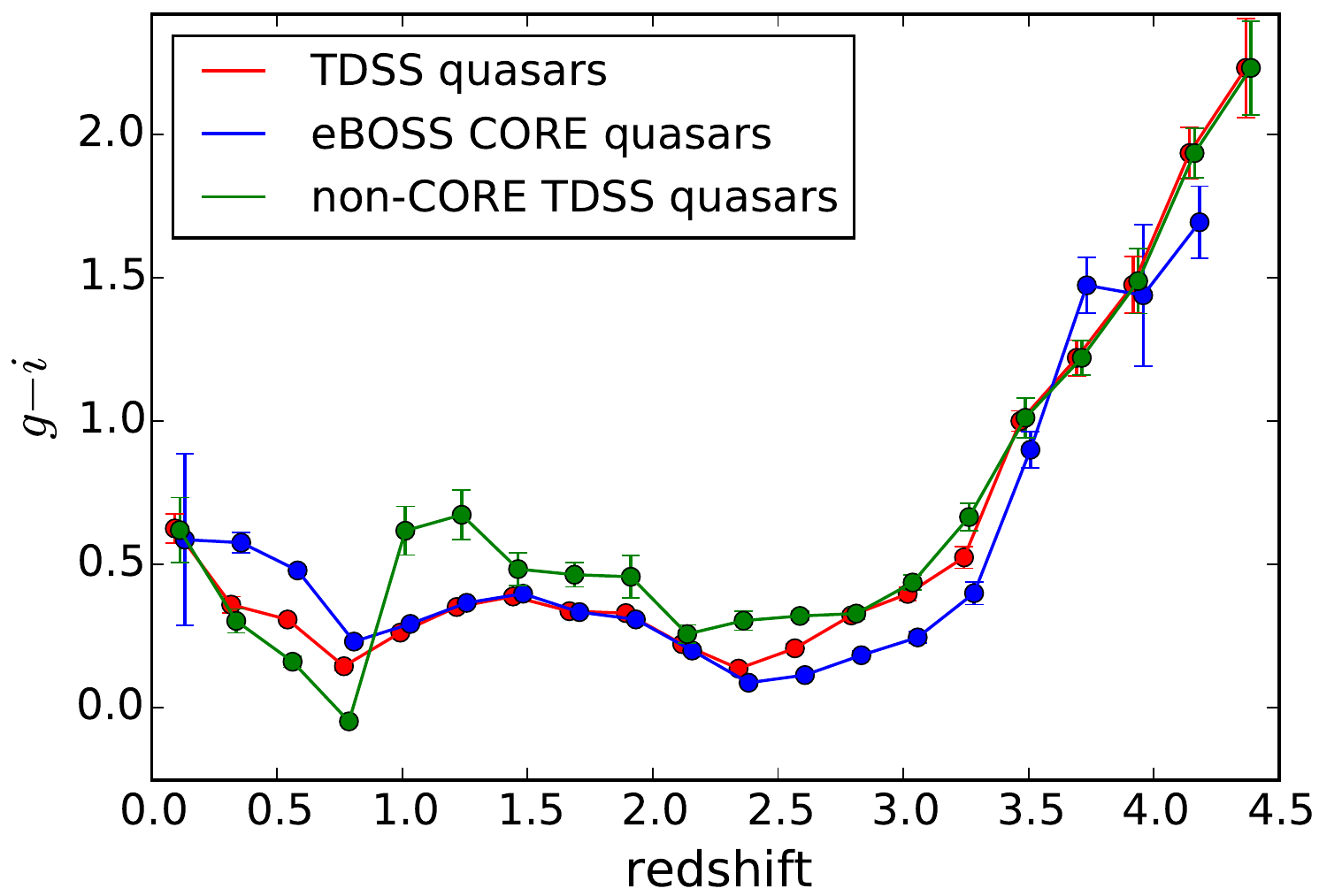}
\caption{Quasar $g-i$ color from SDSS photometry as a function of redshift, for all TDSS-selected (red points), and eBOSS CORE-selected quasars (blue points) with spectra in our SEQUELS plate area. The subset of TDSS-selected quasars not selected by the CORE sample is also shown (green points). TDSS-selected quasars not selected by the eBOSS CORE algorithm in the primary CORE redshift range of $0.9 \lesssim z \lesssim 2.2$ have redder colors, likely due to stronger intrinsic dust extinction or absorption. TDSS quasars at lower- ($z<0.9$) and higher-redshifts ($z>2.2$) have bluer and redder colors relative to the CORE sample, respectively, likely due to selection effects in the color-selected CORE sample.
}
\end{figure}
	
	Figure 5 (top) also shows the normalized redshift distribution for the subset of TDSS-selected quasars in SEQUELS that were not selected by the eBOSS CORE sample. This subset of TDSS-only quasars shows peaks at redshifts $z\sim0.8$ and $z\sim2.7$ (outside of the CORE target redshift range), but also includes additional quasars (with sky density of approximately 1.5 deg$^2$ ) in the $0.9 \lesssim z \lesssim 2.2$ range targeted by the CORE sample; this shows that variability is able to select additional quasars to complement the CORE algorithm in producing a more complete quasar sample. We note that the CORE quasar sample (20,916) is approximately a factor of 2 larger than the TDSS sample (9,925 quasars). To illustrate more clearly the complementary utility of variability-based quasar selection, Figure 5 (bottom) shows the fraction of TDSS-selected quasars in our SEQUELS plate area that were also selected by the eBOSS CORE sample (N$_\mathrm{CORE+TDSS}$/N$_\mathrm{TDSS}$), as a function of redshift. In this comparison, a fraction of N$_\mathrm{CORE+TDSS}$/N$_\mathrm{TDSS}$ = 1 in a redshift bin indicates that all TDSS quasars in that bin were all also independently selected by the CORE algorithm, while a fraction of N$_\mathrm{CORE+TDSS}$/N$_\mathrm{TDSS}$ = 0 indicates that none of the TDSS quasars in that bin were also selected by the CORE algorithm. Over the $0.9 \lesssim z \lesssim 2.2$ redshift range in which eBOSS will attempt to make CORE quasar clustering measurements (and where color selection is highly efficient and complete), the CORE XDQSO algorithm selects a very impressive 94\% of the TDSS quasars. On the other hand, Figure 5 also shows that variability-only selection (i.e. TDSS quasars that were selected by TDSS but not CORE) accounts for more than 40-50\% of TDSS quasars in some redshift bins in this particular comparison, reaffirming the complementary utility of variability-selection. Over the entire redshift range encompassed, variability selection adds about 16\% (by number) to the quasars that would not have been selected by eBOSS CORE color-selection alone \citep[and among variability-only selections, there is additional overlap between TDSS and eBOSS selections using Palomar Transient Factory variability data;][]{palanque15}. The additional quasars selected by TDSS at high redshifts contribute at least in part to the redder colors of TDSS quasars in comparison to eBOSS CORE quasars seen in Figure 4.
	
\begin{figure}[t!]
\centering
\includegraphics[width=0.47\textwidth]{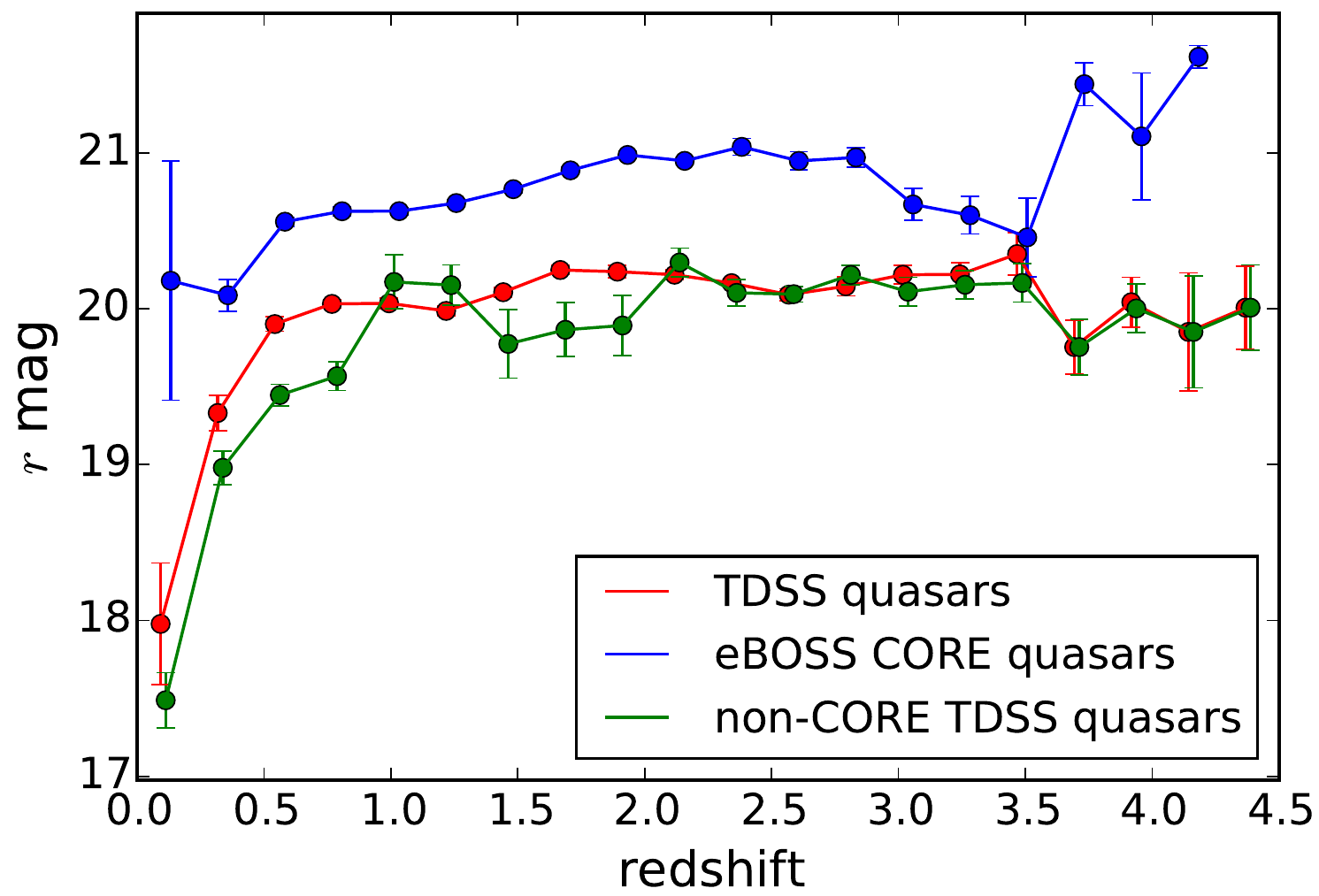}
\caption{Similar to Figure 6, but for observed SDSS $r$-band magnitudes. TDSS-selected quasars are generally brighter than eBOSS CORE quasars, primarily a consequence of the requirement in the TDSS targeting method of robust detections of variability above the photometric uncertainties.
}
\end{figure}

\begin{figure*}[t]
\centering
\includegraphics[width=0.49\textwidth]{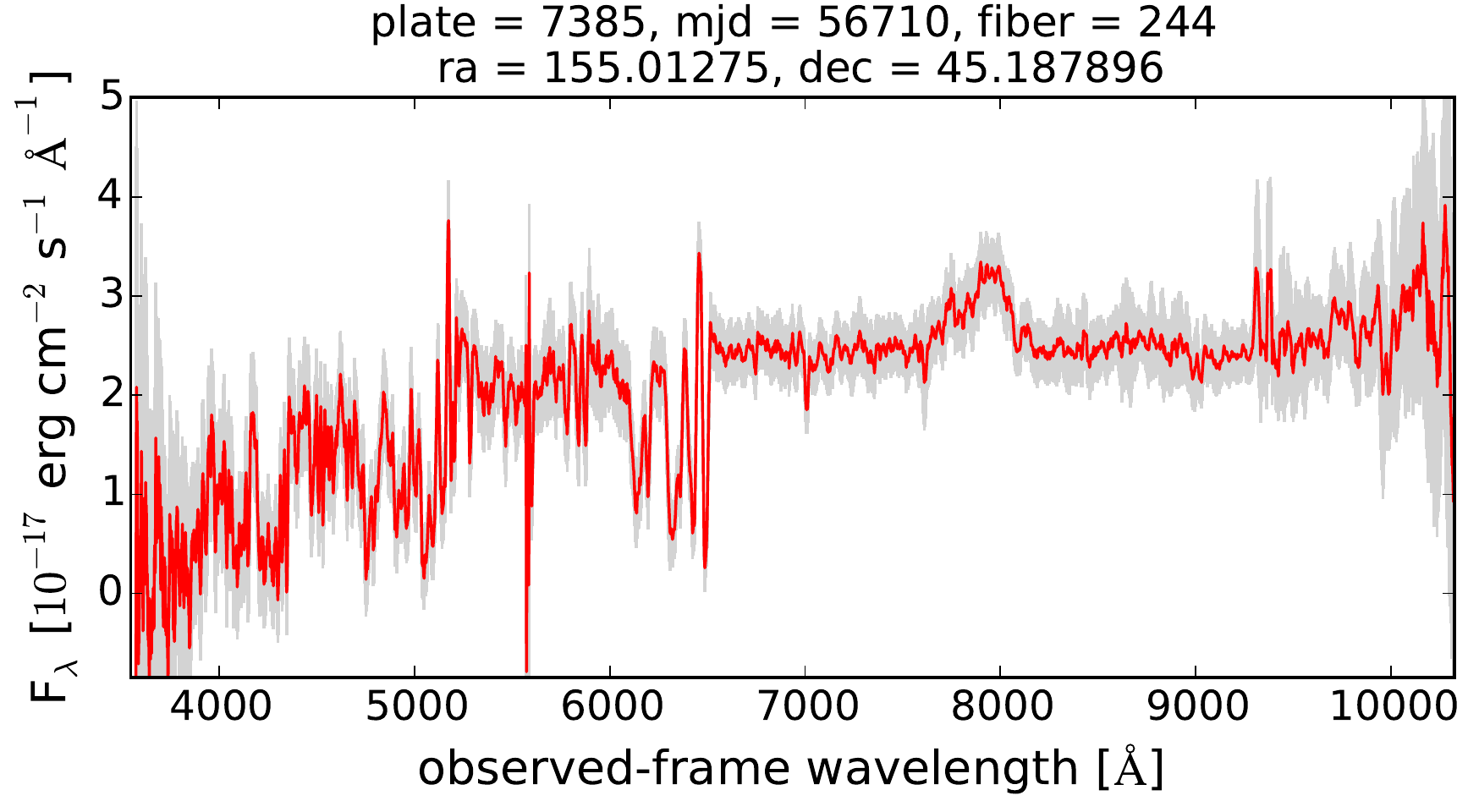}
\includegraphics[width=0.49\textwidth]{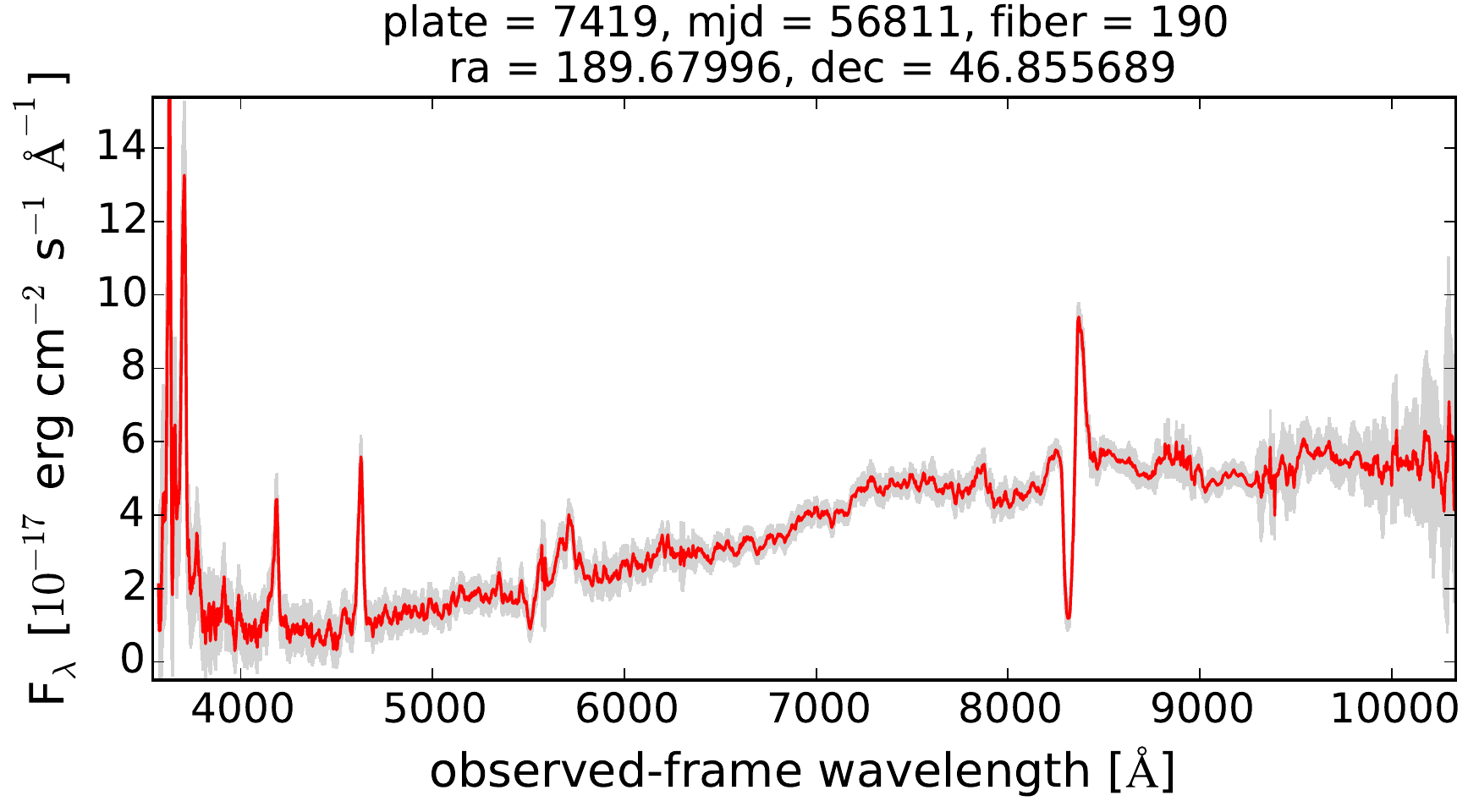}\\
\includegraphics[width=0.49\textwidth]{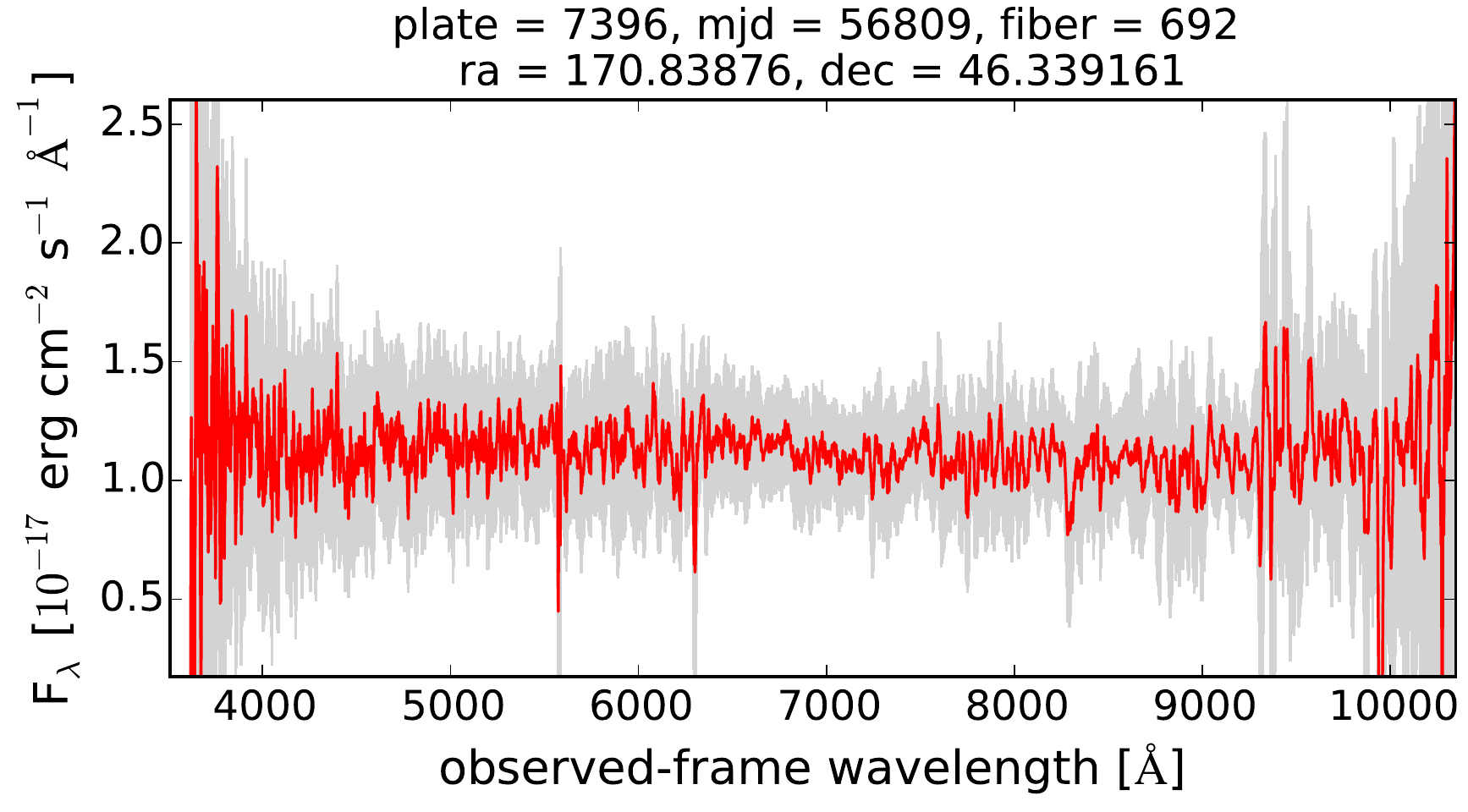}
\includegraphics[width=0.49\textwidth]{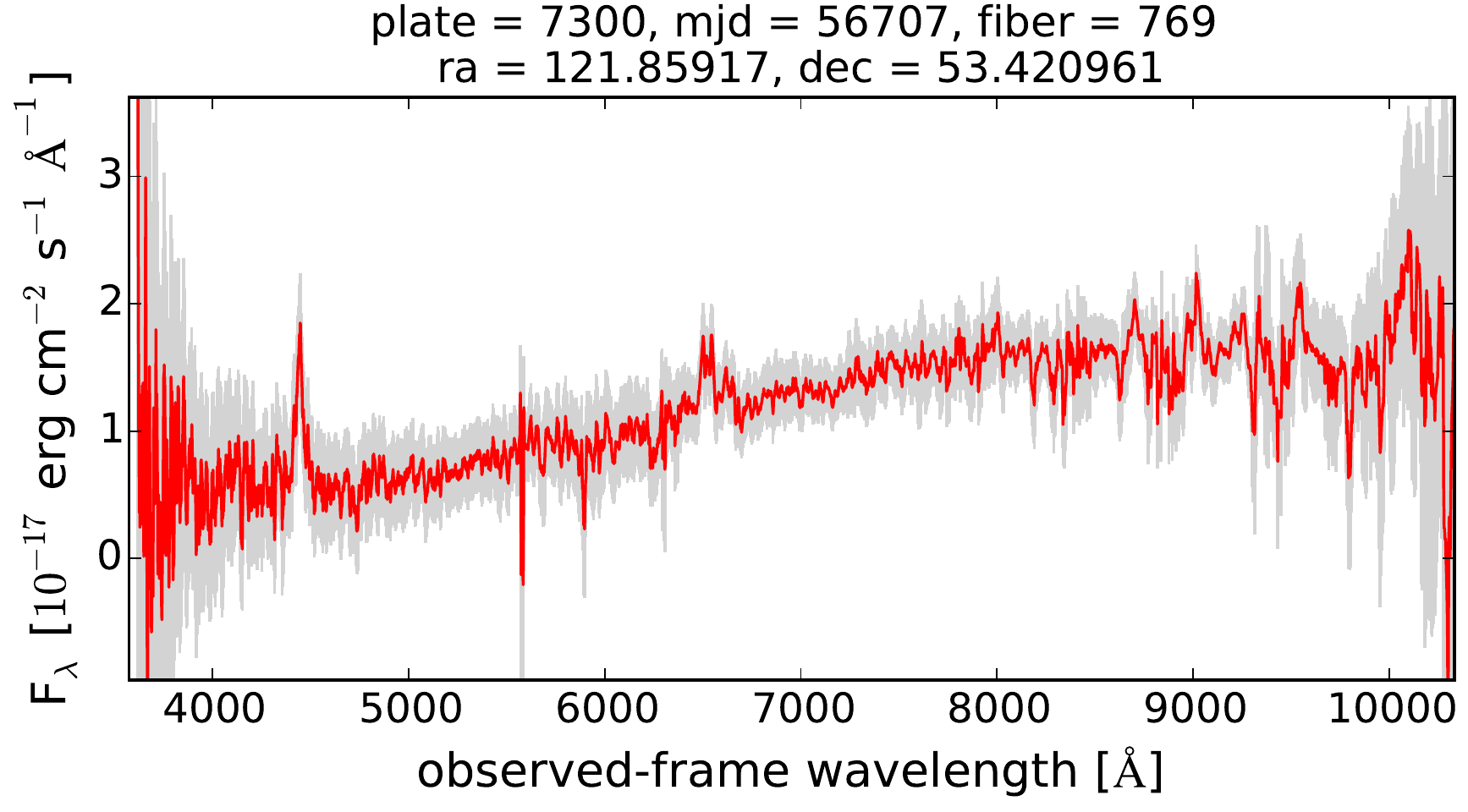}
\caption{Examples of interesting quasar spectra selected only by TDSS in SEQUELS: BAL quasar ($z=3.18$, top left),  LoBAL quasar ($z=1.99$, top right),  blazar candidate with FIRST radio counterpart ($z=0.55$, bottom left), and quasar with red continuum emission ($z=1.33$, bottom right). The spectra are shown in red, and 1$\sigma$ uncertainties in grey.
}
\end{figure*}

\subsection{Quasar Properties with Redshift}
	Quasars with redder, stellar-like colors such as those seen in the TDSS-selected sample in Figure 4 can be in part explained if they are drawn from a population of higher-redshift quasars, but they can alternatively have intrinsically redder colors, e.g.\ due to continuum dust extinction in the host galaxy or substantial intrinsic line absorption. Dust extinction should not affect the continuum variability, while the redder colors can cause these objects to be misclassified as probable stars in color-based selection of quasars; quasars selected on the basis of optical variability may thus be redder than those selected by colors. Luminous quasars with heavy extinction are now commonly found based on their strong infrared emission relative to the optical \citep[e.g.][]{glikman13, ross14}, X-ray emission \citep[e.g.,][]{dwelly06, tozzi06}, or strong narrow emission lines with weak continuum emission \citep[e.g.][]{alexandroff13}. The extinction in these type II quasars is significantly larger than many of the red TDSS-selected broad-line quasars, and they are thus too faint for TDSS to realistically target.
	
	Since the broadband colors of quasars (e.g.\ in Figure 4) are affected by redshift effects, we show in Figure 6 the median colors of these quasar samples as a function of redshift. The 1$\sigma$ uncertainty on the quasar color in each redshift bin is estimated through bootstrap resampling. Over the redshift range of $0.9 \lesssim z \lesssim 2.2$ targeted by the eBOSS CORE sample, the additional quasars selected by TDSS but not the CORE algorithm are systematically redder. The redder colors of these quasars are likely to be due to stronger dust extinction in their host galaxies. At higher redshifts of $z > 2.2$, the redder colors of TDSS quasars relative to the CORE sample may also be due to their intrinsically redder colors. However, an additional effect that can cause TDSS to select redder quasars at $z > 2.2$ is that the CORE algorithm can confuse bluer quasars at these higher redshifts as lying in their targeted $0.9 \lesssim z \lesssim 2.2$ redshift range. This causes the CORE algorithm to select some bluer quasars at $z > 2.2$, which preferentially leaves redder high-redshift quasars to be selected only by TDSS. The redder quasars at $z>0.9$ selected by TDSS but not the CORE algorithm comprise the red quasar clump centered at approximately $g-r$, $u-g$ = [0.2, 1.0] in Figure 4 (right), which has a long tail extending to redder colors.

	Conversely, Figure 6 also shows that TDSS-selected quasars at low redshifts ($z\lesssim0.9$) are instead systematically bluer than CORE quasars. This is likely to be due to the opposite of the effect described above, such that the CORE algorithm is confusing redder quasars at $z\lesssim0.9$ as lying in the $0.9 \lesssim z \lesssim 2.2$ targeted redshift range, thus preferentially leaving bluer low-redshift quasars to be uniquely selected by TDSS. These bluer low-redshift TDSS quasars are the blue quasar clump centered at approximately $g-r$, $u-g$ = [0.0, 0.4] in Figure 4 (right) and discussed in Section 4. Overall, these results show that variability selection can find additional quasars that complement some color-selection approaches. Figure 7 shows the observed SDSS $r$-band magnitude of the three quasar samples, also as a function of redshift. Since TDSS target selection requires robust detection of variability in the light curves above the photometric uncertainties, TDSS quasars are preferentially brighter (by $\sim$1 $r$ mag) than CORE quasars.

\subsection{Unusual Quasars}

	Many unusual quasars are identified in the visual inspection, but have not been rigorously classified.  A few examples of unusual quasars with new spectra in SEQUELS that are selected by TDSS but not by the eBOSS CORE algorithm are shown in Figure 8. We highlight below two major groups of these quasars that are likely to be significant in TDSS: broad absorption line quasars and blazars.

\subsubsection{Broad Absorption Line Quasars}
	Intrinsic broad absorption lines (BALs) are observed in approximately 10-15\% of quasar spectra of sufficient redshift range, almost always in high-ionization lines such as C IV, but also occasionally in low-ionization lines \citep[e.g.,][]{weymann91, hall02, trump06, gibson08}. These BALs have blueshifts of up to $\sim$0.1$c$, indicative of outflowing winds launched from the accretion disk \citep[e.g.,][]{murray95, proga00}, which can affect the properties of the host galaxy \citep{fabian12}. Spectroscopic monitoring of BAL quasars has shown that BAL troughs can vary greatly in depth \citep[e.g.,][]{filiz13} and even disappear \citep[e.g.,][]{filiz12} over timescales of years, and BAL variability has been observed on timescales as short as days \citep{grier15}. It may thus be expected that BAL variability could contribute to increased photometric variability in light curves of BAL quasars, leading to elevated BAL fractions in variability-selected quasar samples relative to color-selected samples. Furthermore, although BAL quasars show a wide range of physical properties, their UV/optical continuum emission is well-known to be systematically redder relative to similar non-BAL quasars \citep[e.g.,][]{reichard03, gibson09}. Since we showed in Section 4 that TDSS-selected quasars have redder continuum emission on average and are less biased in redshift, this may also cause variability-selected samples to contain a higher fraction of BAL quasars relative to color-selected samples, and perhaps one that is actually closer to the true intrinsic BAL fraction. We calculate and compare the BAL fractions between the TDSS and CORE quasar samples in our SEQUELS plate area for quasars at $z>1.7$. This is performed by matching these two samples to the SDSS DR12 BAL quasar properties catalog (Paris et al. 2016, in preparation). We find that 14.1($\pm$0.5)\% of TDSS-selected quasars in SEQUELS are robust BAL cases, in contrast to 9.8($\pm$0.3)\% of CORE-selected quasars. The occurrence of BALs in quasars is thus 43\% higher in the variability-selected TDSS sample than the color-selected eBOSS CORE sample, and is consistent with some previous inferences of the intrinsic BAL fraction \citep{knigge08}.
	
\begin{figure}[t]
\centering
\includegraphics[width=0.47\textwidth]{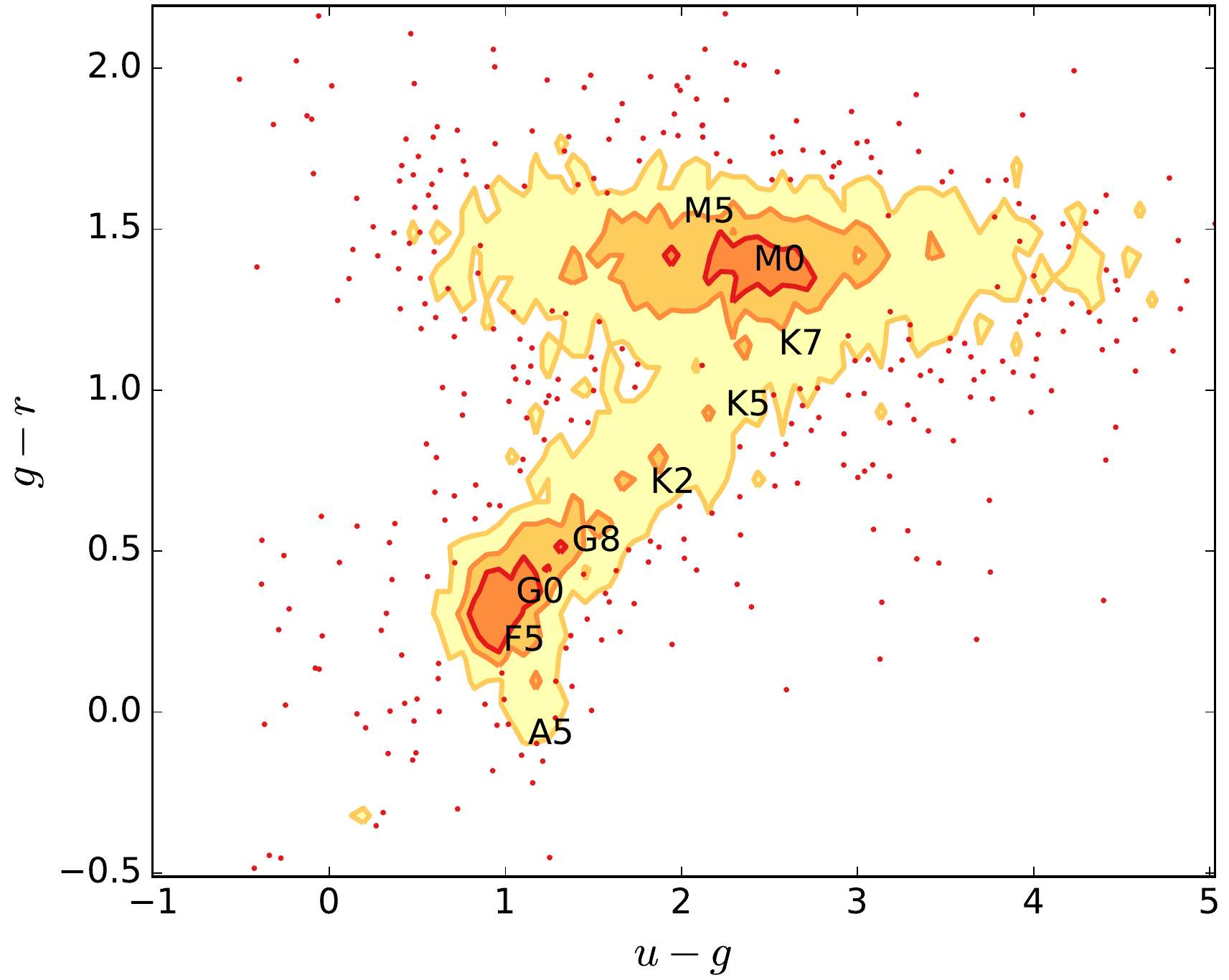}
\caption{Color-color diagram of all TDSS-selected stars with spectra in our SEQUELS plate area (including previous SDSS spectra). The approximate locations of different stellar sub-classes in color-space are labeled based on \citet{kraus07}. The TDSS variable stars sample contains a large number of M dwarfs, A/F stars (many of which may be RR Lyrae), and a significant number of binary stars along the stellar locus. 
}
\end{figure}

\subsubsection{Blazars}
 	Relativistic jets in active galaxies are not uncommon, although only a small fraction are aimed directly along the line of sight of the observer \citep[e.g.,][]{antonucci93, urry95}. These rare jet-beamed objects are classified as blazars, and strong jet synchrotron continuum emission can dominate their optical/UV spectrum. Their optical spectra often appear as featureless power laws with weak/no broad emission lines relative to common unbeamed Type 1 quasars; this may be due to either their lack of broad line regions \citep[in the case of the BL Lac blazar subclass, e.g.][]{nicastro03, elitzur09}, or swamping of the broad emission lines by the jet continuum \citep[in the case of the flat-spectrum radio quasar subclass;][]{giommi12, ruan14b, delia15}. Furthermore, the relativistic Doppler-boosted jet continuum emission is also highly variable, causing blazars to be systematically more variable than non-beamed quasars observed in synoptic surveys \citep{bauer09, ruan12}. Due to this effect, the TDSS sample of quasars is likely to have a significant blazar fraction.

\begin{figure}[t]
\centering
\includegraphics[width=0.47\textwidth]{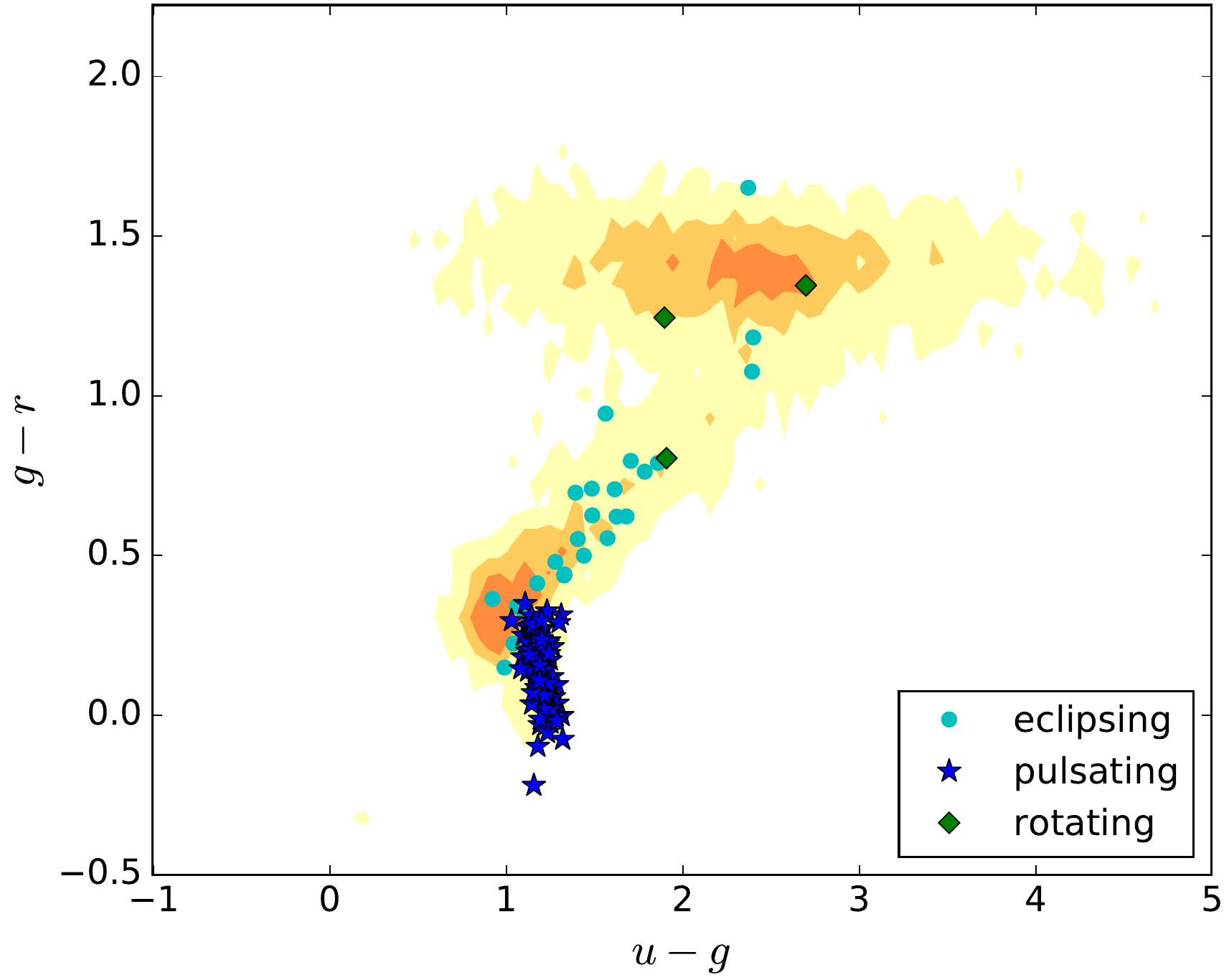}
\caption{Color-color diagram of TDSS stars matched to periodic variable stars from the Catalina survey catalog. The stars are grouped as eclipsing, pulsating, or rotating based on their Catalina classification. The colors of the overall TDSS stellar sample from Figure 9 are shown as shaded contours in the background. We note that this matched sample contains only previously-known bright stars with strong variability, and it is likely there are significantly more periodic stars in the TDSS sample than shown here.
}
\end{figure}
	
\begin{figure*}[t]
\centering
\includegraphics[width=0.33\textwidth]{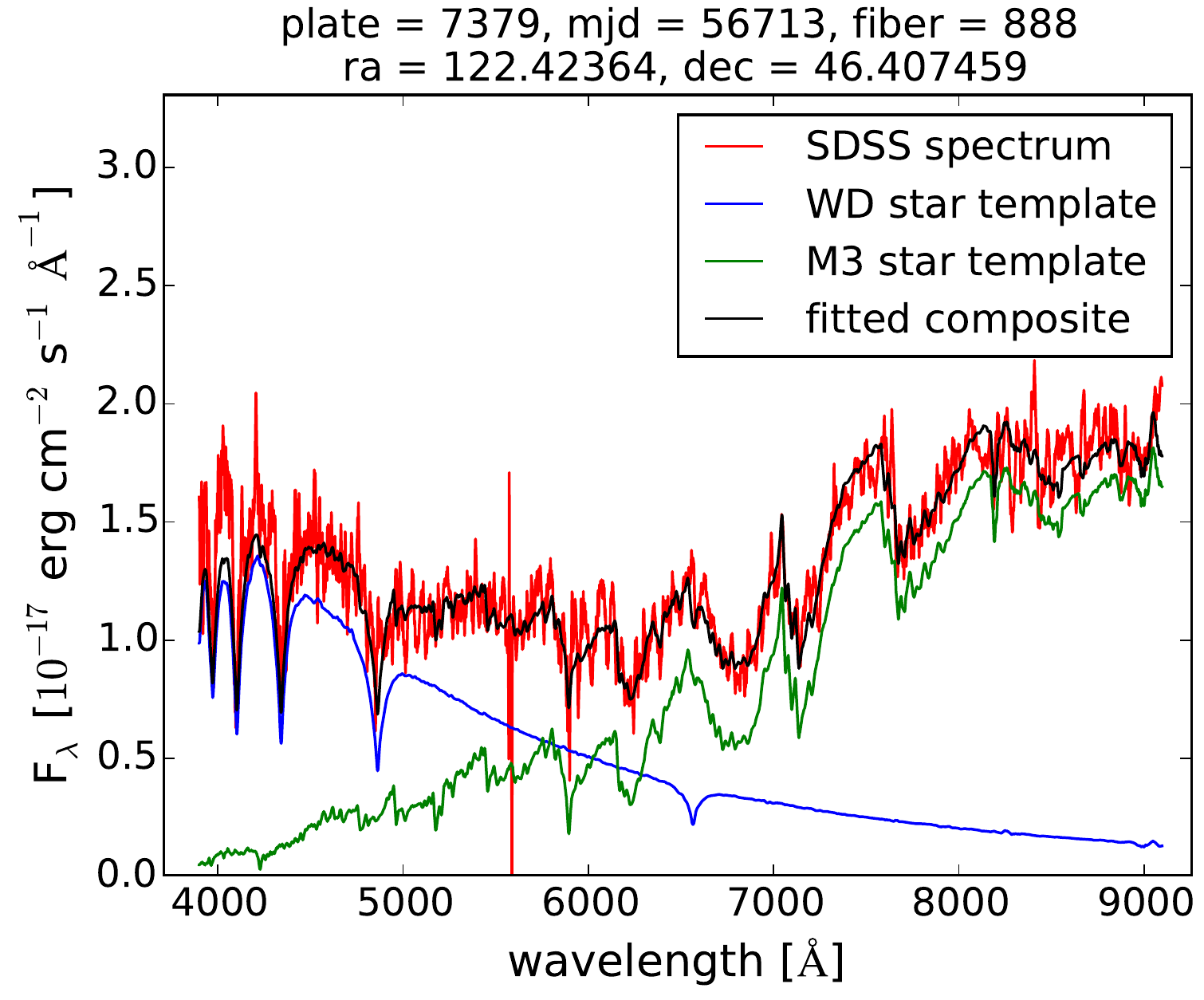}
\includegraphics[width=0.33\textwidth]{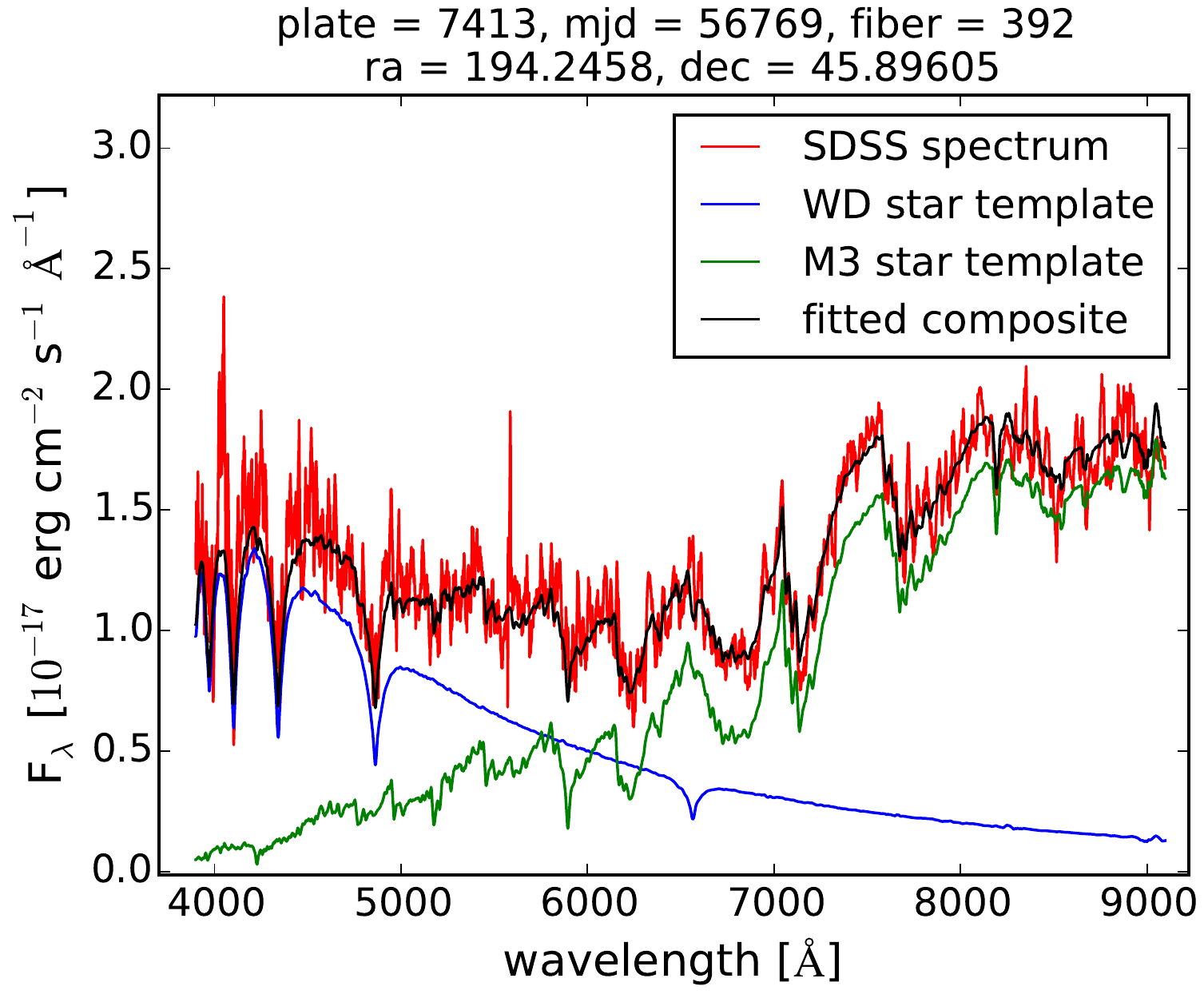}
\includegraphics[width=0.33\textwidth]{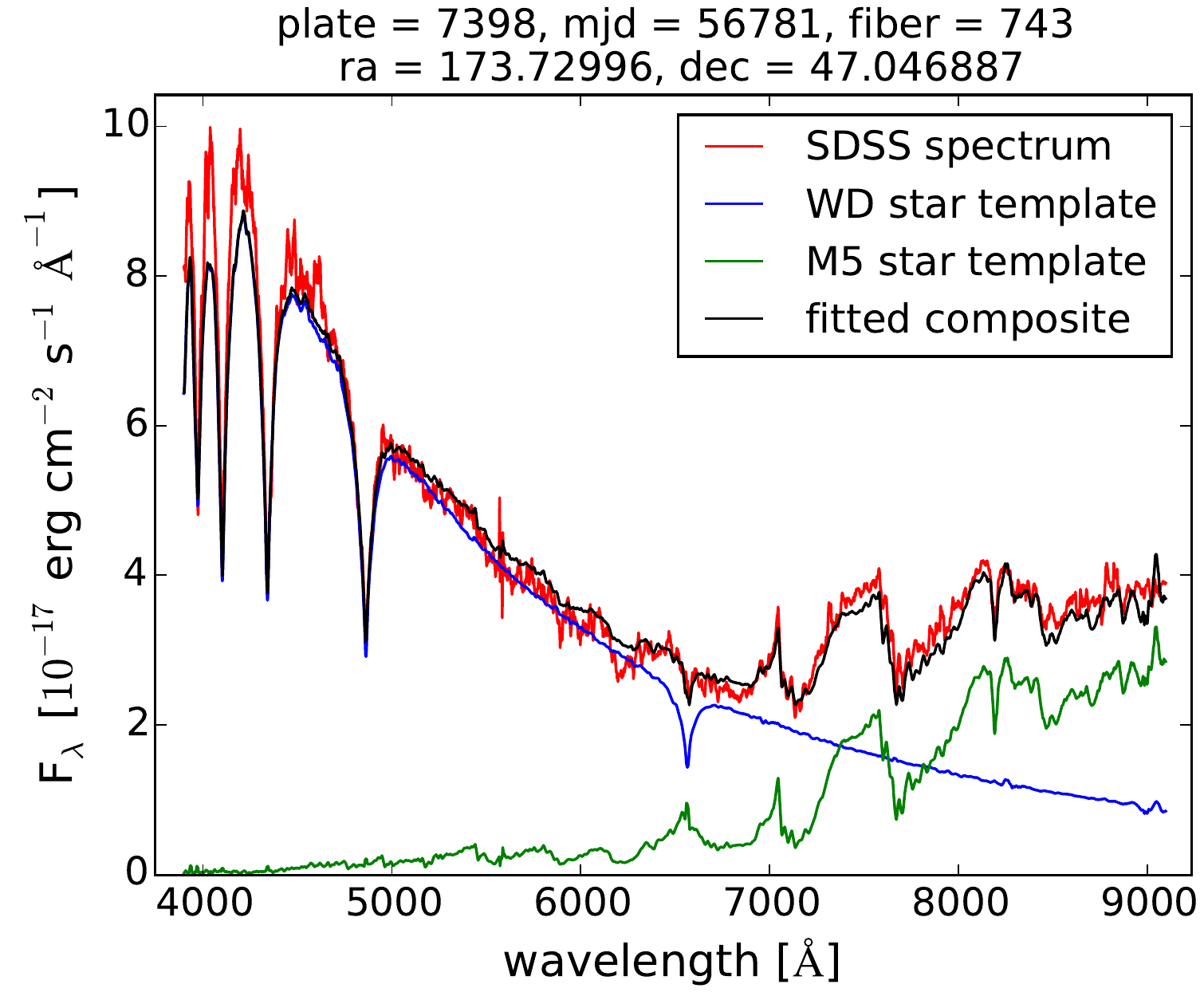}
\caption{Examples of spectral decomposition of a few spectroscopic stellar binaries in the TDSS stars sample identified by visual inspection. Pairs of template spectra from the SDSS-II classification pipeline are fit to the TDSS spectrum through a simple $\chi^2$ fit, and the total composite spectrum provides a good fit to the observed spectra.	
}
\end{figure*}
	
	Blazar candidates in TDSS are identified in our visual inspection based on qualitative assessments of their broad emission line strengths; objects with strong power law-like continuum emission with weak broad emission lines are classified as blazar candidates. Blazars are well-known to almost always be strong emitters in the radio and often in the $\gamma$-rays, in strong contrast to the small $\sim$10\% of unbeamed quasars that are radio loud \citep{ivezic02, balokovic12} and are typically undetected in $\gamma$-rays. A detection of any of our quasars in both of these wavelength regions provides strong support for their blazar nature. We thus match our quasars to the \emph{Fermi} Large Area Telescope \citep{atwood09} 4-year 3FGL catalog of $\gamma$-ray point sources \citep{fermi15}, using the 95\% $\gamma$-ray source position error radii, as well as the FIRST radio survey \citep{becker95}, using a 2\arcsec~matching radius. We find that 26 of our visually-identified quasars are $\gamma$-ray detected, of which 6 are also radio detected (and thus highly likely to be $\gamma$-ray blazars). Our visual inspection also identified 27 BL Lac candidates with characteristic featureless power-law continuum emission, 21 of which are radio-detected in the FIRST survey. This is larger than the 16 high-confidence BL Lac candidates identified similarly by their SDSS spectrum and FIRST radio emission by \citet{plotkin10} in the SDSS Stripe 82 (with sky coverage of approximately 290 deg$^2$, similar to our SEQUELS plate area). Thus, we estimate that our variability-selected sample contains approximately a 30\% higher incidence of BL Lacs than a comparable sample not selected by variability.
	
\section{Stars in TDSS} 

	Although the fraction of stars detected in time-domain imaging surveys that are variable is significantly smaller than that of active galaxies in the magnitude range of TDSS targets, these stars display a wider variety of variability patterns, such as periodic eclipses, atmospheric pulsations, microlensing events, etc. Table 2 shows the approximate distribution in spectral type of TDSS stars in our SEQUELS plate area identified in our visual inspection, and Figure 9 shows a color-color diagram of these stars. The full stellar locus is populated in Figure 9, although two main clumps are apparent: a redder population with M-type colors, as well as a bluer population with A- and F-type colors. We note that the large spread in the $u - g$ direction for the M star clump is primarily due to the larger uncertainties in the $u$-band magnitudes for these red stars rather than being astrophysical in origin. In this section, we will investigate these different variable stellar populations in more detail, concluding that many of the variable stars in the bluer clump in Figure 9 are likely to be pulsating RR Lyrae, the redder clump is likely to be due to the larger fraction of chromospherically active M dwarfs selected by TDSS, while the rest of the stellar locus is likely to contain a large number of eclipsing binaries.
	
	RR Lyrae stars exhibit periodic variability of $\sim$0.5-2 mag with periods of $\sim$0.2-1 day, with distinctive light curve patterns as a result of opacity effects in their atmospheres that lead to pulsations \citep[e.g.,][]{christy66}. RR Lyrae are especially interesting since their high luminosities and observed period-luminosity relation (modulo metallicity) allows them to be used as standard candles to trace the Galactic halo density. Large numbers of RR Lyrae stars discovered using light curves from time-domain surveys have recently revealed a large number of overdensities and substructures in the Galactic halo \citep[e.g.,][]{keller08, sesar10, drake13a, sesar13}. These halo density inhomogeneities have strong implications for our understanding of the formation and evolution of the Galaxy and its constituent parts \citep[e.g., ][]{bullock05, johnston08, bell08, font11}. Spectroscopic follow-up of these RR Lyrae aids in metallicity determination, providing more accurate distance determinations for their use as standard candles. Additionally, radial velocities from RR Lyrae spectra are useful for probing halo substructures, once their considerable pulsation velocities ($\sim$60 km s$^{-1}$  amplitude peak-to-peak) are accounted for.
	
	M dwarfs often display magnetic activity in their chromospheres which can produce stellar flares of several magnitudes with strong blue continuum emission \citep[e.g.,][]{moffet74, hawley91, osten05, kowalski13}. Chromospherically active M dwarfs exhibit strong H$\alpha$ emission, which can be used as a proxy for activity strength in large spectroscopic samples \citep{west04}. These flares can often be observed in stellar light curves from time-domain imaging surveys similar to those used in TDSS \citep{kowalski09, davenport12, davenport14}. Although they are much less luminous than RR Lyrae, the large numbers of M dwarfs makes them useful probes of the structure and kinematics of the thick and thin Galactic disks in the Solar neighborhood \citep[e.g.][]{reid95, hawley96, reid97, bochanski07}. Large samples of M dwarfs have been compiled for these studies, typically based on their high proper motions \citep[e.g.,][]{lepine11}. However, these high proper motion selected samples are known to suffer from kinematic biases, which can be mitigated in part through variability-based selection, especially for later spectral types.
		
	Binary stellar systems can exhibit periodic eclipses in their light curves if the orbital plane is close to the line of sight. The stars in these binary systems are ostensibly formed together, making them interesting laboratories of stellar evolution. A combination of photometric and spectroscopic information can constrain a wide range of binary orbital parameters \citep[e.g.,][]{andersen91, torres10}, although further follow-up in addition to our single-epoch TDSS spectra and SDSS/PS1 light curves will be required for useful constraints in most cases. A significant fraction of TDSS-selected variable stars on the stellar locus in Figure 9 are likely to be a wide variety of eclipsing binary systems. Just a small fraction will show `composite' spectra where both stars in the system are apparent in the TDSS spectra.

\subsection{Periodic Stellar Variables}
	A large fraction of our TDSS-selected stars are likely to be objects that show periodic variability in their light curves. Although the SDSS/PS1 light curves we use may be too sparsely sampled to detect this periodicity, we can assess if some of our TDSS selected objects are known periodically variable stars detected in other surveys. A comparison of the colors of these previously-known periodic variable stars to our TDSS stars sample provides another guide to the different types of stars resulting from the variability-based selection of TDSS. To this end, we positionally match all TDSS-selected objects with spectra in our SEQUELS plate area to the Catalina Sky Survey (CSS) catalog of periodic stars \citep{drake14}, and the CSS RR Lyrae catalogs of \citet{drake13a} and \citet{drake13b}, all with a 2$\arcsec$ matching radius.  Although these catalogs do not extend as faint as our TDSS survey, their higher cadence sampling provides detailed classifications of each periodic star based on their light curve characteristics. Using their detailed classification, we group the matched periodic stars into `eclipsing', `pulsating', and `rotating' subclasses following \citet{drake14} to give a general idea of their variability mechanism. 
			
	Figure 10 shows a color-color diagram of all TDSS-selected stellar objects with spectra in our SEQUELS plate area that matched to the various periodic star catalogs. These stars are dominated by blue RR Lyrae, but span the majority of the stellar locus. This suggests that many of the TDSS-selected blue A- and F-type stars in Figure 9 are likely to be RR lyrae, which is not surprising given their large variability amplitudes and short periods. Furthermore, Figure 10 suggests that the stellar locus of TDSS-selected periodic stars in Figure 9 is likely to be dominated by eclipsing binaries of various types, along with many rotating stellar variables. We caution that these various periodic star catalogs are highly incomplete even for bright objects due to a host of selection effects, and thus here we do not attempt to investigate the expected periodic star populations (and sub-type populations) in our TDSS sample.
	
\begin{figure}[t!]
\centering
\includegraphics[width=0.49\textwidth]{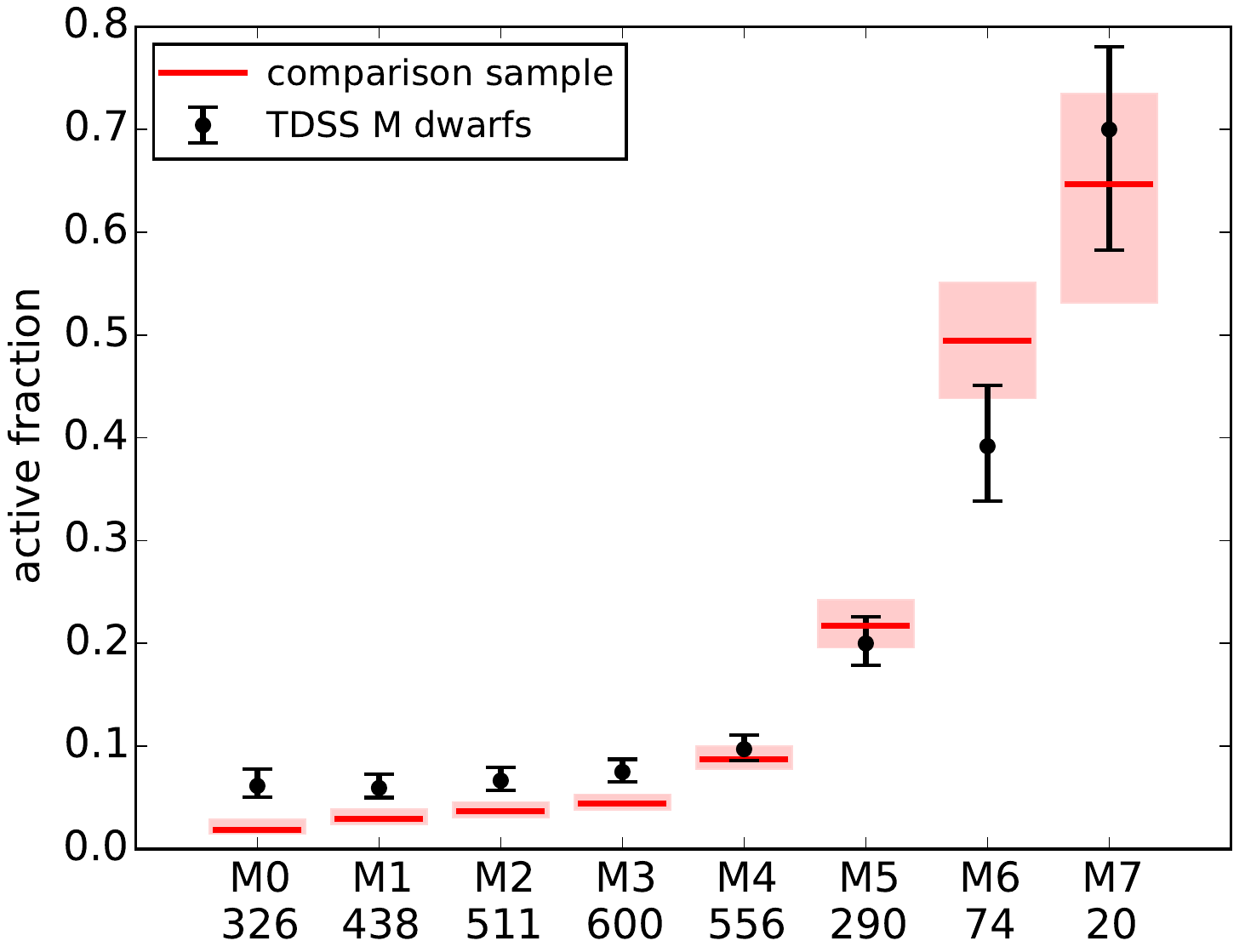}
\caption{Chromospherically active fraction of TDSS M dwarfs for different spectral subtypes, as measured by their H$\alpha$ emission (black points). The number of M dwarfs in each spectral type subsample is also shown. In comparison to the expected active fraction of a sample with the same height-above-Galactic-plane distribution (red bars), early-type TDSS M dwarfs are overall more likely to be active, especially for earlier types. Across all spectral types, the full TDSS M dwarf sample has a 10.0\% overall active fraction, in comparison to the 8.0\% of the comparison sample.
}
\end{figure}
	
\subsubsection{Spectroscopic Composites}
	In our visual inspection, we identified several candidate composite spectra of stars, which are likely to be relatively rare spectroscopic binary systems in which both stars in the system are clearly visible in the optical spectrum. These rare but astrophysically important objects can be identified in the TDSS sample thanks to their variability, commonly due to their eclipsing nature. We spectroscopically decompose the binaries by fitting pairs of template stellar spectra to the observed spectrum to more accurately assess the constituent components of the composite spectrum. Specifically, we utilize the spectral template library from the SDSS-II cross-correlation redshifts pipeline\footnote{http://classic.sdss.org/dr7/algorithms/spectemplates}, and generate synthetic composite spectra using combinations of pairs of these template spectra, through a grid of normalizations. The best-fitting pair of spectra (and the normalizations) are found by minimizing the $\chi^2$ between the synthetic composite spectra and each observed composite spectrum. Figure 11 shows examples of spectra of TDSS dM/WD binary systems (including two rare objects with strong dM emission), along with the results of this simple spectroscopic binary decomposition method. These composite spectrum dM/WD binaries are likely to be rare short-period eclipsing systems, and are selected by TDSS due to their variability.
	
\begin{deluxetable}{cc}
\tablecolumns{12}
\tablewidth{0pt}
\tablecaption{TDSS SEQUELS stellar sample}
\tablehead{\colhead{Spectral type} &  \colhead{Number of spectra}}
\startdata
O/B & 22  \\
A/F & 917  \\
G/K & 1811  \\
M & 2312 \\ 
White dwarfs & 18 \\ 
Spectral composite & 29 \\
Carbon star & 3 \\
Cataclysmic Variable & 14
\enddata

\end{deluxetable}

\begin{figure*}[t]
\centering
\includegraphics[width=0.49\textwidth]{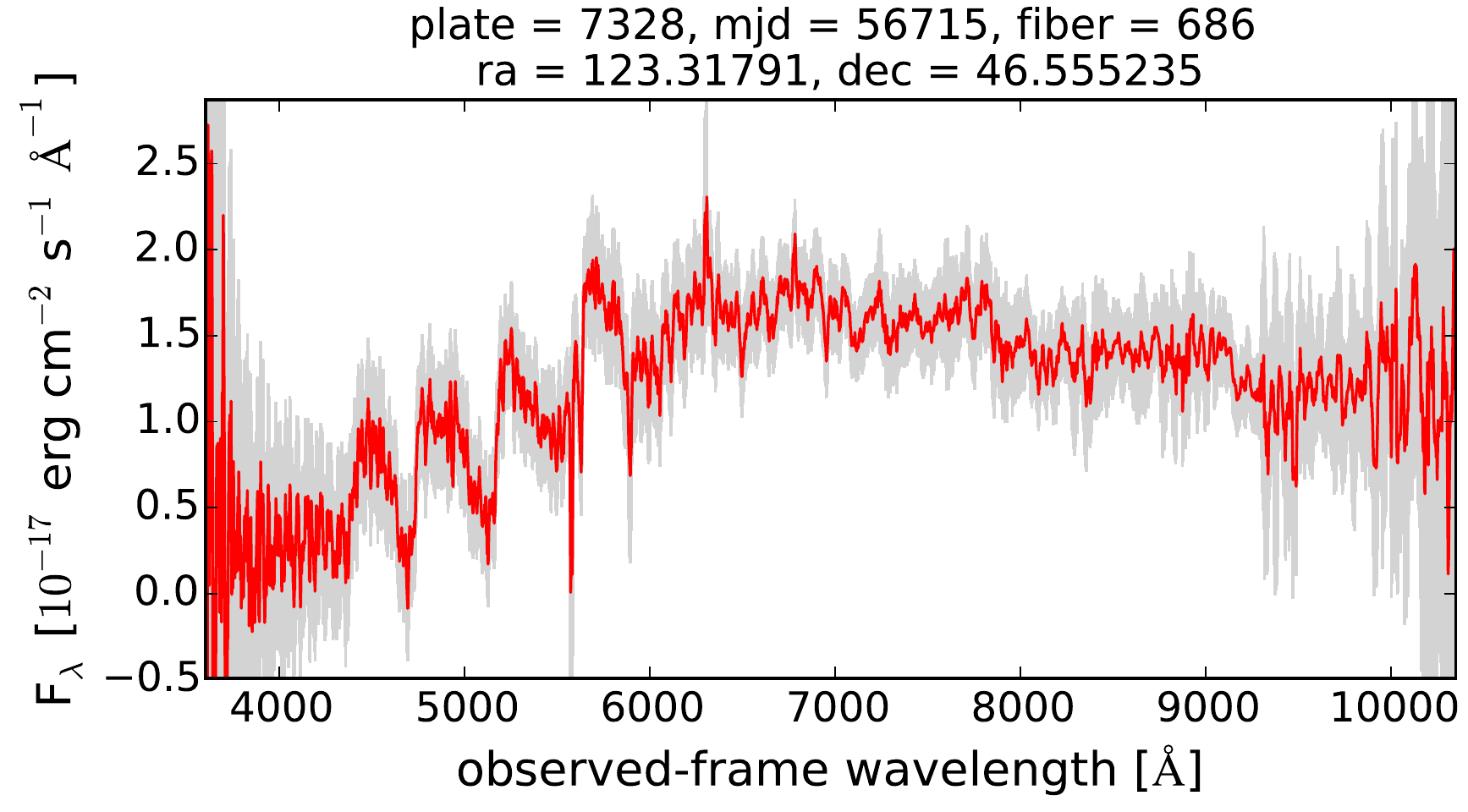}
\includegraphics[width=0.49\textwidth]{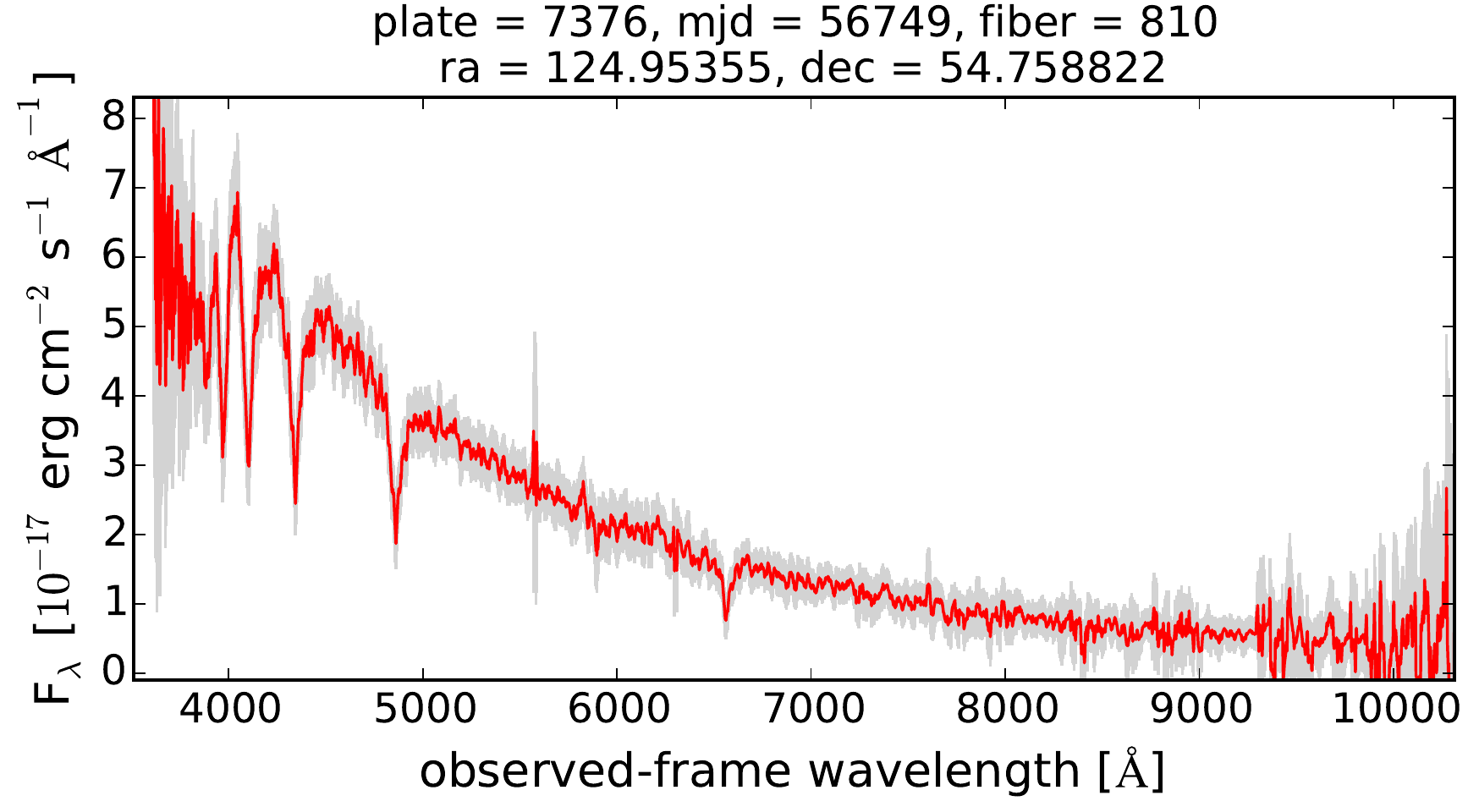}\\
\includegraphics[width=0.49\textwidth]{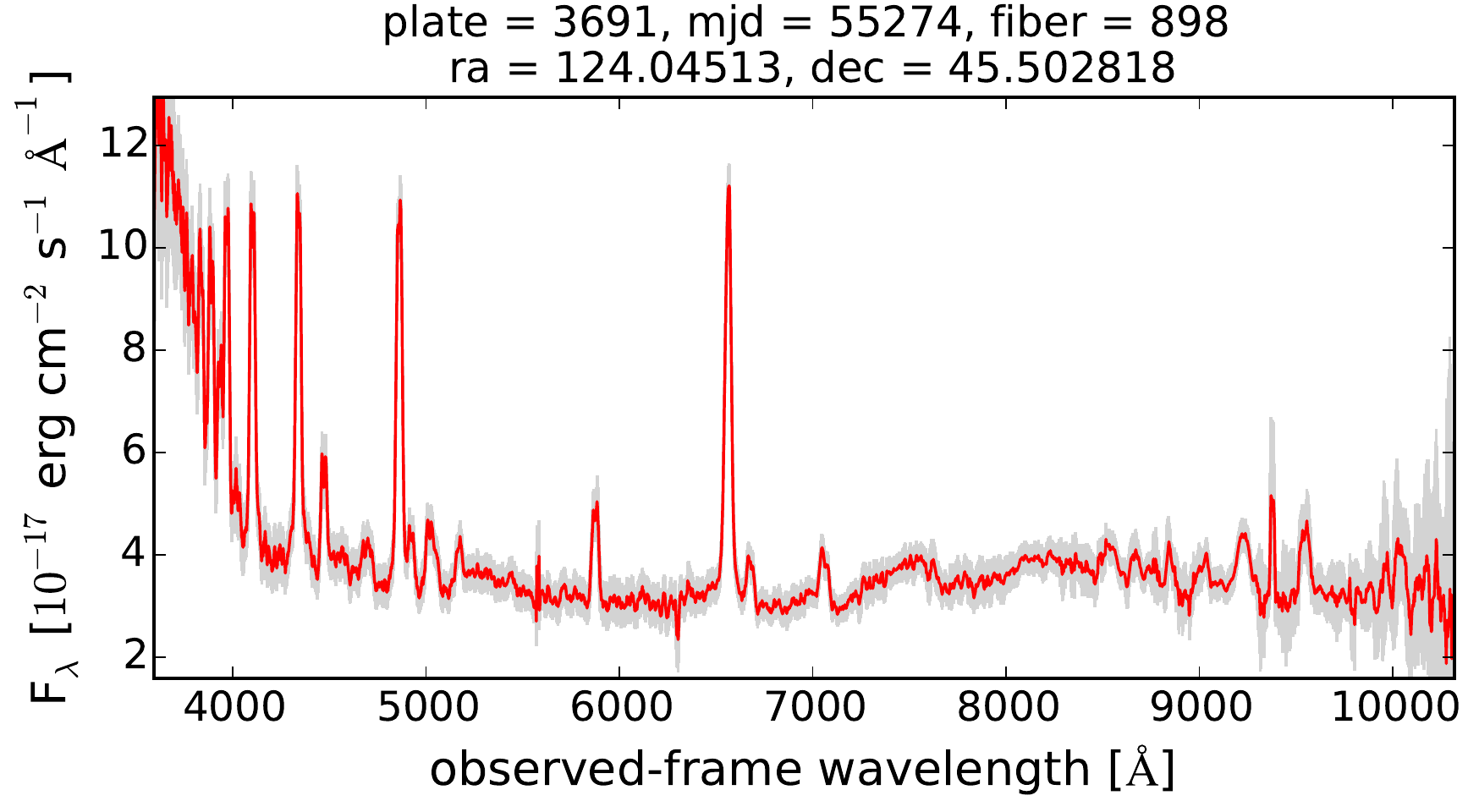}
\includegraphics[width=0.49\textwidth]{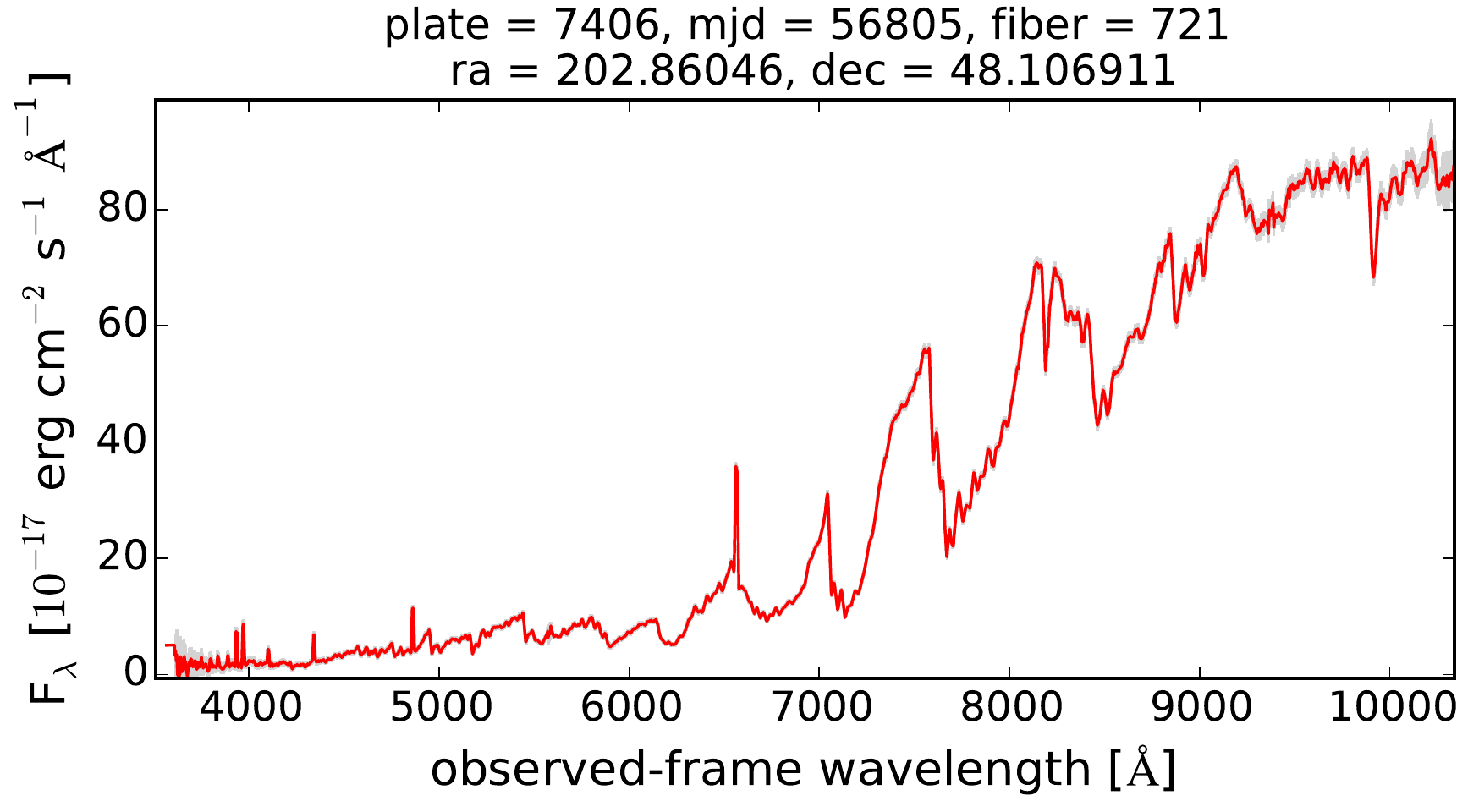}
\caption{Examples of interesting stellar spectra: carbon star (top left), white dwarf (top right), cataclysmic variable (bottom left), chromospherically active M dwarf with H$\alpha$ emission (bottom right). The spectra are shown in red, and 1$\sigma$ uncertainties in grey.
}
\end{figure*}

\subsection{Chromospherically Active Stellar Variables}
	We investigate the chromospherically active fraction of M dwarfs selected by TDSS to assess whether variability yields a higher active fraction than expected from a non-variability selected sample. Chromospheric activity in M dwarfs is well-known to increase for later spectral type M dwarfs up to type M9 \citep{hawley96, schmidt15}, but also decreases with stellar age \citep{gizis02}. In a Galactic context, M dwarfs will show decreasing active fractions for increasing height above the Galactic plane \citep{west06, west08}, since these M dwarfs at large plane heights are on average older due to dynamical heating in the Galactic disk. To take these effects into account, we will compare our TDSS M dwarf active fraction to an estimate of the expected active fraction for a M dwarf sample with the same Galactic plane height distribution, as a function of spectral type.
	 	
	Using all TDSS M dwarfs with spectra in our SEQUELS plate area identified through our visual inspection, we first perform detailed classification using a modified version of the \texttt{Hammer} spectral typing software \citep{covey07}. \texttt{Hammer} allows us to interactively classify each M dwarf through visual comparison to a suite of template spectra. Furthermore, \texttt{Hammer} also selects a subsample of M dwarfs with high signal-to-noise in the H$\alpha$ region for measurements of their H$\alpha$ equivalent width (EW). Specifically, the signal-to-noise ratio of the flux densities in the H$\alpha$ wavelength window and the surrounding continuum wavelength window must be $>$3, which is critical to ensure that our activity classification of each M dwarf as either `active' or `inactive' based on its H$\alpha$ EW is robust and not significantly affected by noise. Following the classification method of \citet{west11}, M dwarfs in our sample are classified as `active' if they have H$\alpha$ EW $>$0.75, with signal-to-noise ratio of the H$\alpha$ EW measurement $>$3. The active fraction is then $N_\mathrm{active}$/($N_\mathrm{active}$+ $N_\mathrm{inactive}$). Figure 12 shows this TDSS M dwarf active fraction as a function of spectral type, with uncertainties calculated from the binomial distribution using the method of \citet{cameron11}. As expected, the active fraction rises for later spectral types.
		
	To compare our TDSS M dwarf active fractions as a function of spectral type to that expected from a non-variability selected sample, we utilize the SDSS DR7 M dwarf catalog of \citet{west11} as a general sample of M dwarfs not specifically based on their photometric variability. To control for difference in height above the Galactic plane (i.e. stellar age) between these two samples, we first determine the height above the Galactic plane for each M dwarf in these two samples. This is done using the color-based photometric distance $M_r$, $r-z$ relation of \citet{bochanski10} with extinction corrected magnitudes. Simultaneously binning the DR7 M dwarfs in height above the Galactic plane and spectral type provides an expectation of the active fraction as a function of these two properties. For each TDSS M dwarf in our sample, we compute this expectation of the active fraction based on its height above the Galactic plane and spectral type. Using these expectations for all of our individual TDSS M dwarfs, the overall expected active fraction of each spectral type is simply the mean expected active fraction for all TDSS M dwarfs, for that spectral type. This results in the active fraction as a function of spectral type of a comparison sample of M dwarfs of the same size that is not variability selected. This is also shown in Figure 12, where the uncertainties on these expected active fractions are also estimated from the binomial distribution. 
	
	Comparison of the active fraction of our TDSS M dwarf sample to that expected from a non-variability selected sample in Figure 12 shows that variability selection selects a higher fraction of active M dwarfs for spectral types M0 through M4. For later types, the active fraction from variability selection may be actually lower than expected, but is more ambiguous due to larger uncertainties. This may hint at a fundamental change in the magnetic dynamo in the interiors of these M dwarfs, as they transition with spectral type from a radiative core with a convective envelope to being completely convective for spectral types later than M4 \citep{west09, reiners10}. The overall active fraction across all spectral types for TDSS is 10.0$^{+0.6}_{-0.5}$\%, in comparison to 8.0$^{+0.5}_{-0.5}$\% for the expected fraction for a sample not selected on the basis of variability, showing that our variability-selected M dwarf sample contains approximately 25\% more active M dwarfs than a sample not selected by variability.
	
\subsection{Unusual Stellar Systems}
	Our visual inspection of the TDSS stellar sample in SEQUELS yielded many spectra of peculiar and/or rare objects, a few examples of which are shown in Figure 13. We defer detailed investigations of these peculiar objects, but note that many of them, including cataclysmic variables, pulsating white dwarfs, and carbon stars, are known to include photometric variables and thus likely to be preferentially selected by TDSS.
	
\begin{figure}[t]
\centering
\includegraphics[width=0.49\textwidth]{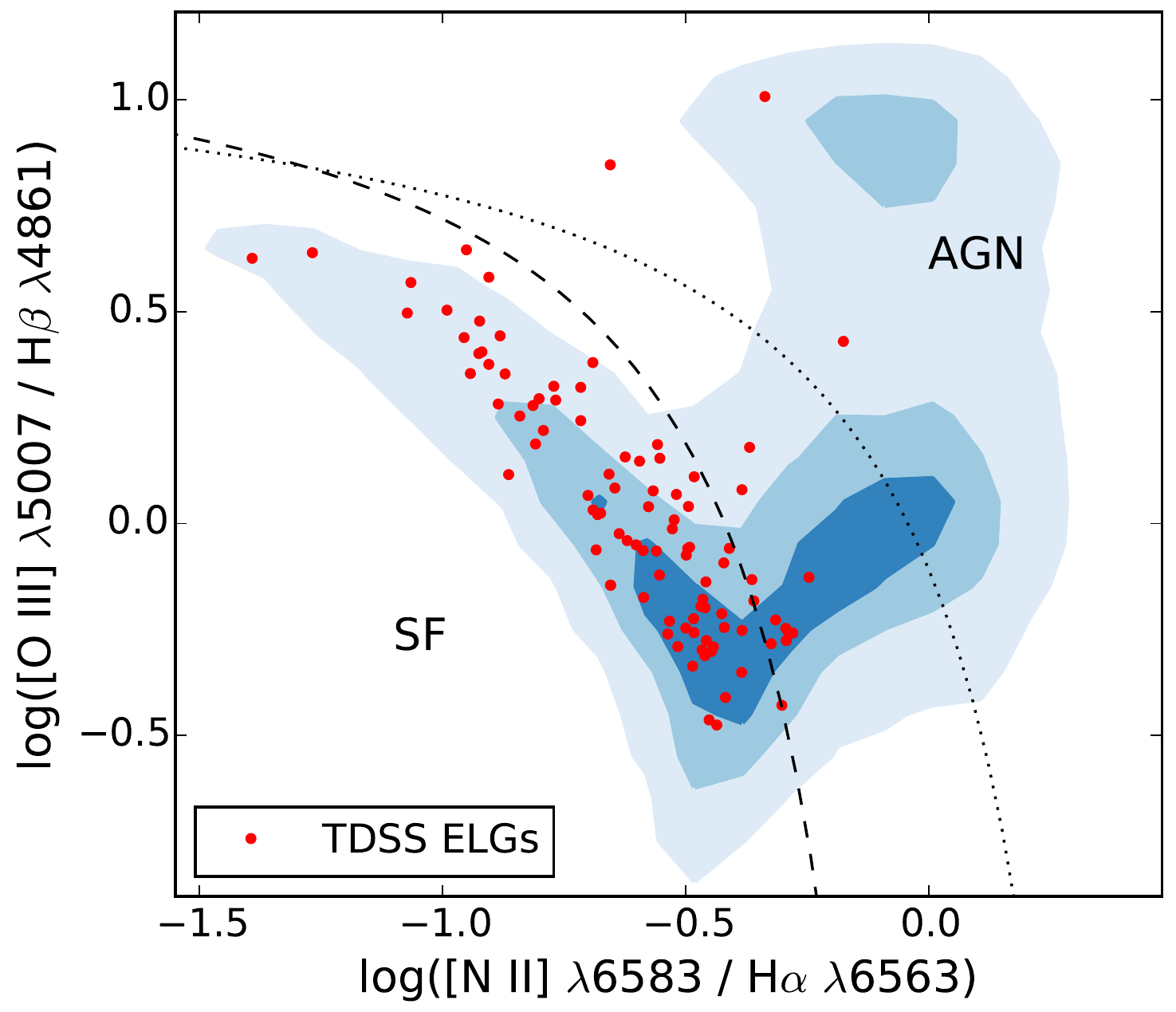}
\caption{BPT diagram of TDSS-selected emission-line galaxies in our SEQUELS plate area (red points), in comparison to all ELGs in DR12 (blue contours). The BPT classifications of \citet{kauffmann03} and \citet{kewley01} are shown as dashed and dotted lines, respectively, marking the nominal divide between star-formation and AGN as the primary source of ionizing flux in these galaxies. The majority of TDSS galaxies do not appear to host active nuclei, and other explanations for their apparent variability should be considered (including possible complications in the photometry of marginally resolved objects).
}
\end{figure}

\section{Galaxies in TDSS} 
	Although TDSS specifically targeted objects classified as point sources in SDSS imaging as part of the targeting algorithm, a small number (313) of galaxies were nevertheless selected and are in our SEQUELS TDSS spectra. These galaxies may be variable due to `hidden' active nuclei, transient phenomena such as tidal disruption events and supernovae, or problems with the photometry leading to artificial variability in the light curves. To investigate whether TDSS-selected galaxies host hidden active nuclei, we will assess the emission line ratios of TDSS emission line galaxies on a classic BPT diagram \citep{baldwin81, kewley01, kauffmann03}, which in effect examines the spectral index of the continuum emission ionizing the emission lines.
		
	To construct the BPT diagram, we match our sample of all TDSS-selected galaxies with spectra in our SEQUELS plate area to the publicly available emission line spectra measurements for all DR12 galaxies by \citet{thomas13}. Specifically, we use the measured [N II]/H$\alpha$ to [O III]/H$\beta$ line ratios for all galaxies with all four lines detected at $>$3$\sigma$ significance. Figure 14 shows this BPT diagram for our TDSS emission line galaxies, compared to all SDSS galaxies in DR12. We also show the BPT classification scheme of \citet{kauffmann03} and \citet{kewley01}, which distinguishes galaxies with emission lines ionized by stellar- and AGN-like continuum spectra. The majority of TDSS galaxies do not appear to host active nuclei (at least at the time of TDSS spectroscopy), and other explanations for their apparent variability should be considered.

	Although other astrophysical explanations are possible, we speculate that many of these galaxies may appear among TDSS selections due to photometry issues, leading to artificial variability in the light curves. In particular, since TDSS targets are selected to be point sources based on their SDSS imaging and passed a round of visual inspection of their imaging before spectra are taken, these galaxies are not obviously extended sources and may instead be barely-resolved. This can lead to small differences in the PSF magnitudes measured by the SDSS and PS1 photometric pipelines, or variations in included flux as seeing changes, leading to artificial variability. However, we note that transient phenomena such as supernovae and stellar tidal disruption events, or `changing look' quasars \citep{lamassa14} in which the AGN continuum disappeared (thus leaving quiescent galaxies in the spectroscopic epoch) may also explain some of these cases. Specifically, because the SDSS spectra are generally observed at a later epoch than the photometric epochs in the light curves (especially for the new spectra in SEQUELS), it is possible that an earlier epoch of strong variability may have abated to result in a quiescent galaxy spectrum.

\section{Conclusions}
	TDSS provides an unprecedented spectroscopic sample of objects selected purely on the basis on their optical flux-variability, useful for a variety of time-domain science to complement light curves from current and future large-scale synoptic imaging surveys. Using an early sample of 15,499 TDSS spectra from the SEQUELS pilot survey for eBOSS, we establish the demographics of our unique variability-selected sample (of which 63\% are quasars and 33\% are stars) through visual inspection of the spectra, providing a spectroscopically-confirmed ground-truth baseline for variability-selected samples. We then investigate the properties of the different classes of objects in our sample, focusing on understanding the unique advantages of using variability-selected samples for specific science applications.
	
	We show that quasars selected by their variability in TDSS display a broad and smooth redshift distribution, complementing color-selected samples by mitigating many color-based redshift biases. For example, in comparison to the color-selected eBOSS CORE quasar sample in the redshift range of $0.9 < z < 2.2$ intended for eBOSS quasar clustering measurements, TDSS selects additional redder quasars. The redder colors of these quasars selected by variability (without regard to color) are possibly due to intrinsic dust extinction or absorption. The TDSS quasar sample also contains an elevated fraction of BAL quasars that is $\sim$40\% higher than the color-selected CORE sample (and likely to be closer to the intrinsic BAL fraction). We also identify other types of peculiar quasars that may be especially common in a variability selected sample, such as blazars which we estimate to be $\sim$30\% more common in TDSS than color-selected samples.
	
	We investigate the stellar sample yielded by TDSS, which spans the full stellar locus but includes a large number of A/F type stars and M type stars. Through comparisons to catalogs of previously-known periodic variable stars, we show that the majority of A/F type stars in our sample are likely to be RR Lyrae, and many of the main-sequence TDSS stars are likely to be eclipsing binary systems. Our large sample of late type stars is likely to contain chromospherically active M dwarfs, and we show that the TDSS M dwarf sample has a higher chromospherically active fraction than expected from a non-variability selected sample, based on their H$\alpha$ emission. We also identify several composite spectra of binary stellar systems in which both stars in the system are clearly visible in the optical spectrum; these rare but astrophysically important systems can be identified in the variability-selected TDSS sample likely due to their eclipsing nature.
	
	Finally, we search for evidence of hidden active nuclei in the small number of emission-line galaxies selected as part of TDSS by studying their narrow emission line ratios in a BPT diagram. We find that the majority of these galaxies do not appear to host active nuclei, and speculate that their appearance in our variability-selected sample may be due to PSF photometry errors if they are barely-resolved (although transient events in these galaxies can also cause their light curves to appear highly variable while producing a galaxy-like spectrum at the later spectroscopic epoch).

	The TDSS survey is the first spectroscopic study of a large sample of objects selected using inclusive variability criteria, and our investigation here lays the groundwork for more in-depth studies of particular classes of objects for a diverse variety of science cases. An additional major science goal of TDSS is the discovery of rare objects often thought to exist but heretofore unobserved (`known unknowns'), as well as completely new classes of objects (`unknown unknowns'). Discovery of these rare objects requires robust outlier detection in large samples; we plan on continued visual inspection of the $\sim$220,000 spectra in the main TDSS sample that will be observed as part of SDSS-IV over the period of 2014-2020. Looking forward, the main TDSS spectroscopic sample will provide a powerful training set for identifying and understanding the hundreds of millions of variable objects detected in future large imaging surveys, and the combination of these light curves and future planned multi-object spectrographs can be exploited to more fully realize the rich science potential in time-domain astronomy.

\acknowledgments
JJR acknowledges NASA support through Fermi Guest Investigator grant NNX14AQ23G. WNB acknowledges NSF support through grant AST-1516784.

Funding for SDSS-III has been provided by the Alfred P. Sloan Foundation, the Participating Institutions, the National Science Foundation, and the U.S. Department of Energy Office of Science. The SDSS-III web site is http://www.sdss3.org/.

SDSS-III is managed by the Astrophysical Research Consortium for the Participating Institutions of the SDSS-III Collaboration including the University of Arizona, the Brazilian Participation Group, Brookhaven National Laboratory, Carnegie Mellon University, University of Florida, the French Participation Group, the German Participation Group, Harvard University, the Instituto de Astrofisica de Canarias, the Michigan State/Notre Dame/JINA Participation Group, Johns Hopkins University, Lawrence Berkeley National Laboratory, Max Planck Institute for Astrophysics, Max Planck Institute for Extraterrestrial Physics, New Mexico State University, New York University, Ohio State University, Pennsylvania State University, University of Portsmouth, Princeton University, the Spanish Participation Group, University of Tokyo, University of Utah, Vanderbilt University, University of Virginia, University of Washington, and Yale University.

The PS1 Surveys have been made possible through contributions of the Institute for Astronomy, the University of Hawaii, the Pan-STARRS Project Office, the Max-Planck Society, and its participating institutes, the Max Planck Institute for Astronomy, Heidelberg, and the Max Planck Institute for Extraterrestrial Physics, Garching, The Johns Hopkins University, Durham University, the University of Edinburgh, Queen's University Belfast, the Harvard-Smithsonian Center for Astrophysics, and the Las Cumbres Observatory Global Telescope Network, Incorporated, the National Central University of Taiwan, and the National Aeronautics and Space Administration under Grant No. NNX08AR22G issued through the Planetary Science Division of the NASA Science Mission Directorate.

\bibliography{bibref}
\bibliographystyle{apj}
\clearpage

\end{document}